\documentclass[runningheads]{llncs}
\usepackage{amsmath,amssymb}
\usepackage{lscape, rotating}
\usepackage{graphicx}
\usepackage{extarrows}

\newtheorem{rem}{Remark}

\begin{document}

\pagestyle{plain}

\mainmatter

\title{Security Concerns in Minimum Storage Cooperative Regenerating Codes
\protect\footnote{$^1$Kun Huang and Ming Xian are with State Key Laboratory of Complex Electromagnetic Environment Effects on Electronics and Information System, National University of Defense Technology, Changsha, 410073, China \email{(khuangresearch923@gmail.com; qwertmingx@tom.com)}.\\
$^2$ Udaya Parampalli is with Department of Computing and Information Systems, University of Melbourne,  VIC 3010, Australia \email{(udaya@unimelb.edu.au)}.\\
}
}
\titlerunning{}

\author{
Kun Huang\inst{1},\ Udaya Parampalli\inst{2}, \and\ Ming Xian\inst{1}
}

\authorrunning{}

\institute{}

\maketitle

\begin{abstract}

Here, we revisit the problem of exploring the secrecy capacity of minimum storage cooperative regenerating (MSCR) codes under the $\{l_1,l_2\}$-eavesdropper model, where the eavesdropper can observe the data stored on $l_1$ nodes and the repair downloads of an additional $l_2$ nodes. Compared to minimum storage regenerating (MSR) codes which support only single node repairs, MSCR codes allow efficient simultaneous repairs of multiple failed nodes, referred to as a \emph{repair group}. However, the repair data sent from a helper node to another failed node may vary with different repair groups or the sets of helper nodes, which would inevitably leak more data information to the eavesdropper and even render the storage system unable to maintain any data secrecy.

\setlength{\parindent}{2em} In this paper, we introduce and study a special category of MSCR codes, termed ``\emph{stable}" MSCR codes, where the repair data from any one helper node to any one failed node is required to be independent of the repair group or the set of helper nodes. Our main contributions include: 1. Demonstrating that two existing MSCR codes inherently are not stable and thus have poor secrecy capacity, 2. Converting one existing MSCR code to a stable one, which offers better secrecy capacity when compared to the original one, 3. Employing information theoretic analysis to characterize the secrecy capacity of stable MSCR codes in certain situations.

\end{abstract}

\section{INTRODUCTION}
Distributed storage systems (DSSs) are an essential infrastructure for the generation, analysis and archiving of tremendously growing data. DSSs have been becoming a fundamental and indispensable component of those rapidly developing distributed networking applications, especially in cloud computing, social networking and peer to peer networking. In order to guarantee DSSs' reliability and availability, data redundancy has to be introduced. Replication and erasure codes are two traditional approaches to offer data redundancy, while erasure codes can achieve higher reliability for the same level of redundancy when compared to replication \cite{Re:H.Weatherspoon}. Recently, Dimakis et al. \cite{Re:A.Dimakis} employ network information flow to determine a class of regenerating codes, which has superior performance over traditional erasure codes regarding repair efficiency.

\subsection{Regenerating Codes}

Regenerating codes \cite{Re:A.Dimakis} are a family of codes determined by trading off the amount of storage per node with the repair bandwidth. In the regenerating-coding-based DSSs, an original data file of size $B$ is encoded into $n\alpha$ symbols and then distributed across $n$ nodes. These symbols can be drawn from a finite field $\mathbb{F}_q$ and each node stores $\alpha$ symbols. The basic features of regenerating codes are reconstruction and regeneration properties, that is, the original data file can be retrieved by contacting any $k$ out of $n$ nodes and any failed node can be recovered by permitting a new node to connect to any $d$ helper nodes from the remaining $(n-1)$ nodes by downloading $\beta$ symbols from each node. Regenerating codes are always parameterized by $\{n, k, d, \alpha, \beta,B\}$ and have the following constraint (tradeoff curve):
\begin{equation}\label{cut}
B\leq \sum_{i=1}^{k}\min\{\alpha,(d-i+1)\beta\}.
\end{equation}
Most of studies now focus on the two extreme points, famous as minimum storage regenerating (MSR) codes and minimum bandwidth regenerating (MBR) codes. As shown in \cite{Re:A.Dimakis}, the parameters of the two points are given by
\begin{equation}
\left\{\begin{aligned}
&(\alpha_{\mathbf{MSR}},\beta_{\mathbf{MSR}})=(\frac{B}{k}, \frac{B}{k(d-k+1)})\\
&(\alpha_{\mathbf{MBR}},\beta_{\mathbf{MBR}})=(\frac{2dB}{k(2d-k+1)}, \frac{2B}{k(2d-k+1)}).
\end{aligned}\right.
\end{equation}
Besides, there are three repair models considered in the literature: functional repair, exact repair, and exact repair of systematic nodes \cite{Re:A.G.Dimakis1}. In contrast, exact repair is preferred in the practical systems since the lost data in any failed nodes can be regenerated exactly \cite{Re:C.Huang}. In the scenario of exact repair, the authors in \cite{Re:Shah.N.B} demonstrated the nonachievability of most interior points on the storage-bandwidth tradeoff curve. For those interior points that might be achievable, coding construction appears rarely \cite{Re:C. Tian,Re:T. Ernvall}.

So far, there are many explicit constructions with exact repair property. In \cite{Re:K.Rashmi}, the authors utilize product matrix framework to propose MBR codes for all parameters and MSR codes under the constraint $\{d\geq 2k-2\}$. In the MSR scenario, much progress has been made. From a global point of view, there are two main classes of MSR codes, namely the scalar MSR codes with $\{\beta=1\}$ \cite{Re:K.Rashmi,Re:C.Suh,Re:Y.Wu1,Re:K.V.Rashmi,Re:Rashmi,Re:N. B. Shah} and vector MSR codes with $\{\beta=(n-k)^{x}\}$ where $x\geq 1$ \cite{Re:I.Tamo,Re:Z. Wang,Re:Z. Wang1,Re:D. S. Papailiopoulos,Re:V. R. Cadambe,Re:V. R. Cadambe1,Re:G. K. Agarwal}. Most of these constructions are heavily built on the concept of interference alignment. According to the analysis in \cite{Re:N. B. Shah}, interference alignment is the necessity of constructing scalar linear MSR codes and these scalar linear MSR codes only exist when $d\geq 2k-2$. These codes as well correspond to the low rate regime (i.e.,$\frac{k}{n}\leq \frac{1}{2n}+\frac{1}{2}$). For designing the high rate codes with $\{\frac{k}{n}>\frac{1}{2}\}$, the vector MSR codes are applicable as they are free of the parameter constraints $(n,k)$. However, many of these vector codes allow efficient repair of only systematic nodes \cite{Re:I.Tamo,Re:Z. Wang1,Re:V. R. Cadambe,Re:V. R. Cadambe1,Re:G. K. Agarwal}. Technically speaking, those MSR codes restricted to only efficient systematic repair are not formal MSR codes, since the formal ones require that any failed nodes including parity nodes should be efficiently repaired. Given this concern, the authors in \cite{Re:Z. Wang,Re:D. S. Papailiopoulos} present vector MSR codes allowing efficient repair of all nodes in different ways. In addition to the repair efficiency, Zigzag code \cite{Re:I.Tamo} has the optimal update property and optimal access property while its variant \cite{Re:Z. Wang} also has the optimal access property. These two properties are of significant value to practical implementations. Furthermore, locally repairable codes lately have attracted a lot of attention due to its practical performance \cite{Re:N.Prakash,Re:P.Gopalan,Re:D. S. Papailiopoulos1}.
\vspace{0.2cm}

As we know, all the above repair mechanisms are designed for single node failure. However, it is also common that DSSs may experience multiple node failures. Sometimes, DSSs, such as Total Recall \cite{Re:R. Bhagwan}, may take the lazy repair policy, where the repair is triggered only when the number of node failures reaches a default threshold. Although most of the existing regenerating codes can in principle be exploited for handling multiple node failures by sequentially applying multiple single node repair procedures, however, they are not optimal in terms of repair bandwidth as explained in \cite{Re:Y. Hu}.

\subsection{Cooperative Regenerating Codes}

To allow efficient repair of multiple simultaneous node failures and further reduce the total repair overhead, Y. Hu et al. \cite{Re:Y. Hu} propose the cooperative regenerating codes. Different from regenerating codes, the repair process of cooperative regenerating codes is divided into two steps which have to handle $t$ node failures. In the first step, $t$ new nodes connect to any $d$ surviving nodes, where each new node needs to download $\beta$ symbols from each helper node (surviving node). In the second step, these $t$ new nodes switch to a process of cooperative repair by exchanging $\beta'$ symbols with each other, where the exchanging data actually is the function of the repair data obtained from the first repair step. In the terminology, the $t$ new nodes are always called as a repair group. Later, the authors in \cite{Re:A.-M. Kermarrec,Re:K. W. Shum2,Re:F. Oggier} derive the tradeoff curve between storage per node and repair bandwidth for cooperative regenerating codes. Similar to regenerating codes, cooperative regenerating codes achieving the two end points of the trade off curve are termed minimum bandwidth cooperative regenerating (MBCR) code and minimum storage cooperative regenerating (MSCR) code respectively. The corresponding parameter set $\{n, k, d, t, \alpha, \beta, \beta', B\}$ of the two points are
\begin{equation}
\left\{\begin{aligned}
&(\alpha_{\mathbf{MSCR}},\beta_{\mathbf{MSCR}},\beta'_{\mathbf{MSCR}})=(\frac{B}{k}, \frac{B}{k(d-k+t)},\frac{B}{k(d-k+t)})\\
&(\alpha_{\mathbf{MBCR}},\beta_{\mathbf{MBCR}},\beta'_{\mathbf{MBCR}})=(\frac{(2d+t-1)B}{k(2d-k+t)}, \frac{2B}{k(2d-k+t)},\frac{B}{k(2d-k+t)}).
\end{aligned}\right.
\end{equation}
Here, we make a comparison on repair bandwidth between MSR and MSCR codes. Assume that there is a storage system with $\{n,k,d,B\}$ and $t$ is the threshold on the number of failed nodes. For MSR codes, every one of $t$ failed nodes needs to contact any $d$ out of $(n-t)$ surviving nodes and downloads the repair data, which totally produces $\frac{tdB}{k(d-k+1)}$ repair bandwidth. For MSCR codes, recovering all the $t$ failed nodes needs $\frac{t(d+t-1)B}{k(d-k+t)}$ repair bandwidth in total. By contrast, it is apparent that when $t>1$,
\begin{equation}
\frac{t(d+t-1)B}{k(d-k+t)}<\frac{tdB}{k(d-k+1)},
\end{equation}
which exactly means that MSCR codes are advantageous over MSR codes when repairing multiple node failures. However, there are not many constructions of cooperative regenerating codes up to now.

Authors in \cite{Re:K. W. Shum1,Re:S. Jiekak,Re:A. Wang} present explicit constructions of MBCR codes and the code proposed in \cite{Re:A. Wang} is built for all parameter settings. In the MSCR scenario, there are only a few constructions \cite{Re:N. Le Scouarnec,Re:K. W. Shum,Re:J. Li}. The construction in \cite{Re:N. Le Scouarnec} is based on the special parameter settings that $k=t=2$ and the one in \cite{Re:K. W. Shum} is limited to the case $d=k$. In \cite{Re:J. Li}, the authors establish an equivalent connection between exact MSR codes and exact MSCR codes, such that linear scalar exact MSCR codes with $\{n,k,d-1,t=2\}$ can be built from any instance of linear scalar exact MSR codes with $\{n,k,d\}$.

\vspace{0.2cm}

Despite the above crucial issues on node failures in DSSs, there always exist security problems since massive storage nodes are widely spread across the network. Accordingly, it will be preferable to incorporate security requirements during the design of the cooperative-regenerating-coding based DSSs. Our concern in this paper is the data secrecy of MSCR-coding-based DSSs.

\subsection{Secrecy Concerns in DSSs}
The active attacker and passive attacker models are the two usual adversary models considered in the literature \cite{Re:H.Delfs}. For the active adversary model, the attacker can take operations on certain compromised nodes such as modifying, injecting and deleting. In this paper, we focus on the passive adversary model, where an adversary can only eavesdrop the data stored on some $l_1$ nodes and repair downloads for other $l_2$ nodes.

{\bf Related work (secure regenerating codes):} The authors in \cite{Re:S.Pawar} and \cite{Re:N.B.Shah} firstly investigate the problem of designing secure DSSs against eavesdropping. In \cite{Re:S.Pawar}, the authors analyze the secrecy capacity of regenerating codes, based on an initial adversary model where the contents of $l<k$ nodes are eavesdropped. They derive an upper bound of the secrecy capacity and propose a secure MBR coding scheme that can attain this bound:
\begin{equation}\label{initial bound}
B^{(s)}\leq \sum_{i=l+1}^{k}\min\{\alpha,(d-i+1)\beta\}.
\end{equation}
Afterwards, the authors in \cite{Re:N.B.Shah} extend the initial eavesdropping model considered in \cite{Re:S.Pawar}, where the eavesdropper can also observe the repair downloads for additional $l_2$ nodes apart from the data stored on the initial $l_1$ nodes, with the constraint that $l_1+l_2<k$. The secure product-matrix-based MBR coding scheme proposed in \cite{Re:N.B.Shah} is shown to achieve the bound (\ref{initial bound}) only by changing $l$ into $l_1+l_2$. The achievability follows from the fact that the repair bandwidth $d\beta$ is equal to per node storage $\alpha$ in the MBR scenario. Furthermore, the authors in \cite{Re:N.B.Shah} considered designing secure product-matrix-based MSR codes, but the secrecy capacity of their secure MSR coding scheme is only $(k-l_1-l_2)(\alpha-l_2\beta)$, which is evidently less than $(k-l_1-l_2)\alpha$ when $l_2>0$ given in the bound (\ref{initial bound}). The reason is that the amount of repair downloads $d\beta$ is larger than the per node storage $\alpha=(d-k+1)\beta$ and thus the $(l_1,l_2)$-eavesdropper can obtain more information in addition to the contents of $(l_1+l_2)$ nodes in the MSR scenario.

Recently, the authors in \cite{Re:Rawat.A.S} and \cite{Re:S.Goparaju} employ the analysis of linear subspace intersection and then derive new upper bounds on secrecy capacity for MSR codes. Zigzag code \cite{Re:I.Tamo} and its variant \cite{Re:Z. Wang} are shown to achieve these new bounds through pre-coding of maximum rank distance (MRD) code \cite{Re:Gabidulin,Re:R. M. Roth}. The bounds given in  \cite{Re:S.Goparaju} match to those in \cite{Re:Rawat.A.S} when $l_2\leq2$. Thereafter, we \cite{Kun} utilize the information theoretic analysis to give some novel results on the secrecy capacity for MSR codes, which includes some new insights on general MSR codes and provides generalized bounds on secrecy capacity for linear MSR codes. Thereby, we demonstrate that the secure product-matrix-based MSR codes given in \cite{Re:N.B.Shah} are also optimal whenever $l_1+l_2\leq k-1$ and $l_2\leq d-k+1$. The final outcome on secrecy capacity of linear MSR codes that we present in \cite{Kun} exhibits to be closely related to the parameter $\beta$ and applies to all known MSR codes including the scalar MSR codes as well as the vector MSR codes like Zigzag code \cite{Re:I.Tamo}. Moreover, it is also applicable to those unexplored vector MSR codes with parameters $\{1<\beta<d-k+1\}$. Furthermore, we find that all of these results also apply to systematic MSR codes with repair data of systematic nodes captured.

{\bf Related work (secure cooperative regenerating codes):} In \cite{Re:Koyluoglu}, the authors pioneer the research of secrecy capacity of cooperative regenerating codes by min-cut analysis. Similar to MBR codes, the total repair bandwidth of MBCR codes under a repair group is also identical to the total storage of the $t$ failed nodes. Thus, the secrecy capacity of MBCR codes are fully characterized under the $\{l_1,l_2\}$-eavesdropping model. For MSCR codes, they derived some results on secrecy capacity in some special cases and claimed that the two existing MSCR codes \cite{Re:N. Le Scouarnec,Re:K. W. Shum} can be transformed into secure MSCR codes. However, they only considered the information leakage under single repair group and neglected an important detail of the repair property in the MSCR scenario. \footnote{Although a node in different repair groups appears in different repair scenarios and corresponds to distinct newcomer nodes, these distinct newcomer nodes corresponding to the same node must appear separately and cannot exist simultaneously in the storage system. Since in the model, the eavesdropper is defined to be capable of observing the repair downloads of certain nodes at the same time, these newcomer nodes corresponding to the same node that, however, cannot appear simultaneously, thus can be viewed as one node if eavesdropped.}Due to different repair groups involving a node whose repair downloads are eavesdropped, the eavesdropper may obtain different repair data sent from a helper node to this eavesdropped node, which will definitely result in more information leakage. Even worse, it may be impossible for storage system to keep any data secrecy after traversing all possible repair groups. Let us briefly describe it as follows.

Suppose there is an MSCR-coding-based DSS specified by $\{n,k,d,t=2,B\}$ and the repair downloads of node $1$ is observed by the eavesdropper. We let $S^{1_{(1,i)}}_j$ denotes the repair data sent from the surviving node $j$ to the failed node $1$ under the repair group $(1,i)$, where $i\neq j$. However, if storage system successively undergoes two different repair groups $(1,i_1)$ and $(1,i_2)$ where $i_1\neq i_2$ and $S^{1_{(1,i_1)}}_j\neq S^{1_{(1,i_2)}}_j$, the eavesdropper will observe more data information. In the worst case, the eavesdropper may obtain all the original data information only needing to wait for traversing all possible repair groups including node $1$. Thus, it will be difficult or even impossible to retain the data secrecy if this kind of MSCR codes is used.

{\bf Contributions:} In this work, we study the data secrecy issue of MSCR codes under the $\{l_1,l_2\}$-eavesdropper model. Considering the possible impacts on security mentioned above, we introduce a new class of MSCR codes, termed ``stable" MSCR codes, where the repair data is restricted to be independent of repair group and the set of helper nodes. In order to elaborate the importance of this ``stable" property to security, we reanalyze the two existing MSCR codes \cite{Re:N. Le Scouarnec,Re:K. W. Shum}. We demonstrate that they both inherently are not stable. The MSCR code given in \cite{Re:N. Le Scouarnec} actually offers no secrecy at all under the $\{l_1=0,l_2=1\}$-eavesdropper model, which makes it impossible to be transformed into a secure MSCR code. In addition, we find that the other MSCR code given in \cite{Re:K. W. Shum} has poor secrecy capacity, even also losing any data secrecy in some cases. Subsequently, we convert the MSCR code given in \cite{Re:K. W. Shum} to a stable one via adjusting its repair strategy.

Then, we turn to investigate the secrecy capacity of stable MSCR codes. Based on precoding using MRD codes, we give an information theoretic expression of secrecy capacity for general MSCR codes. By studying the basic properties of reconstruction and multiple simultaneous regenerations for general MSCR codes and stable MSCR codes, we derive a series of information theoretic features on the contents of node's storage and the repair downloads. Afterwards, combining these features with the secrecy expression, we present a simple expression of secrecy capacity for stable MSCR codes and some specific characterizations on secrecy capacity. A similar result given in \cite{Re:Koyluoglu} is a special case of ours when $d=k$, while the authors therein only considered under single repair group. Finally, we calculate the specific secrecy capacity of the stable MSCR code built from conversion, which is consistent with our information theoretic results on secrecy capacity and is clearly better than that of the original unstable one.

\subsection{Organization}

Section 2 gives preliminaries about system model and adversary model from information theoretic perspective. Section 3 exhibits the detailed illustration of two existing MSCR codes. Section 4 presents some basic information theoretic properties of general MSCR codes and stable MSCR codes. Section 5 provides main results on secrecy capacity of stable MSCR codes. Section 6 concludes this paper.

\section{PRELIMINARIES}

In this section, we describe the system model and the eavesdropping model from information theoretic perspective. In addition, we give the definition of ``stable" MSCR codes.

\vspace{0.4cm}

{\bf A. \emph{Repair Terminology}:} Consider a DSS consisting of $n$ storage nodes. After $t$ nodes fail, $t$ new nodes are introduced to replace these failed nodes. These $t$ new nodes constitute a \emph{repair group}. Each new node connects to any $d$ same surviving nodes and downloads $\beta$ symbols from each of these $d$ nodes. In the cooperative repair phase, each new node contacts the other $t-1$ new nodes in the same repair group and downloads $\beta'$ symbols from each of these nodes. So, the nodes participating in a failed node's repair can be categorized into surviving nodes (the $d$ helper nodes) and cooperative nodes (the other $t-1$ new nodes). In addition, the repair downloads involved in the system also can be divided into ``repair data" (from the surviving nodes) and ``exchanging data" (from the cooperative nodes). Here, it should be noted that the exchanging data is not necessarily the function of the data stored in the original failed node and actually is the function of the repair data of the corresponding new node.

\begin{figure}[!h]
  \centering
  \includegraphics[width=12 cm]{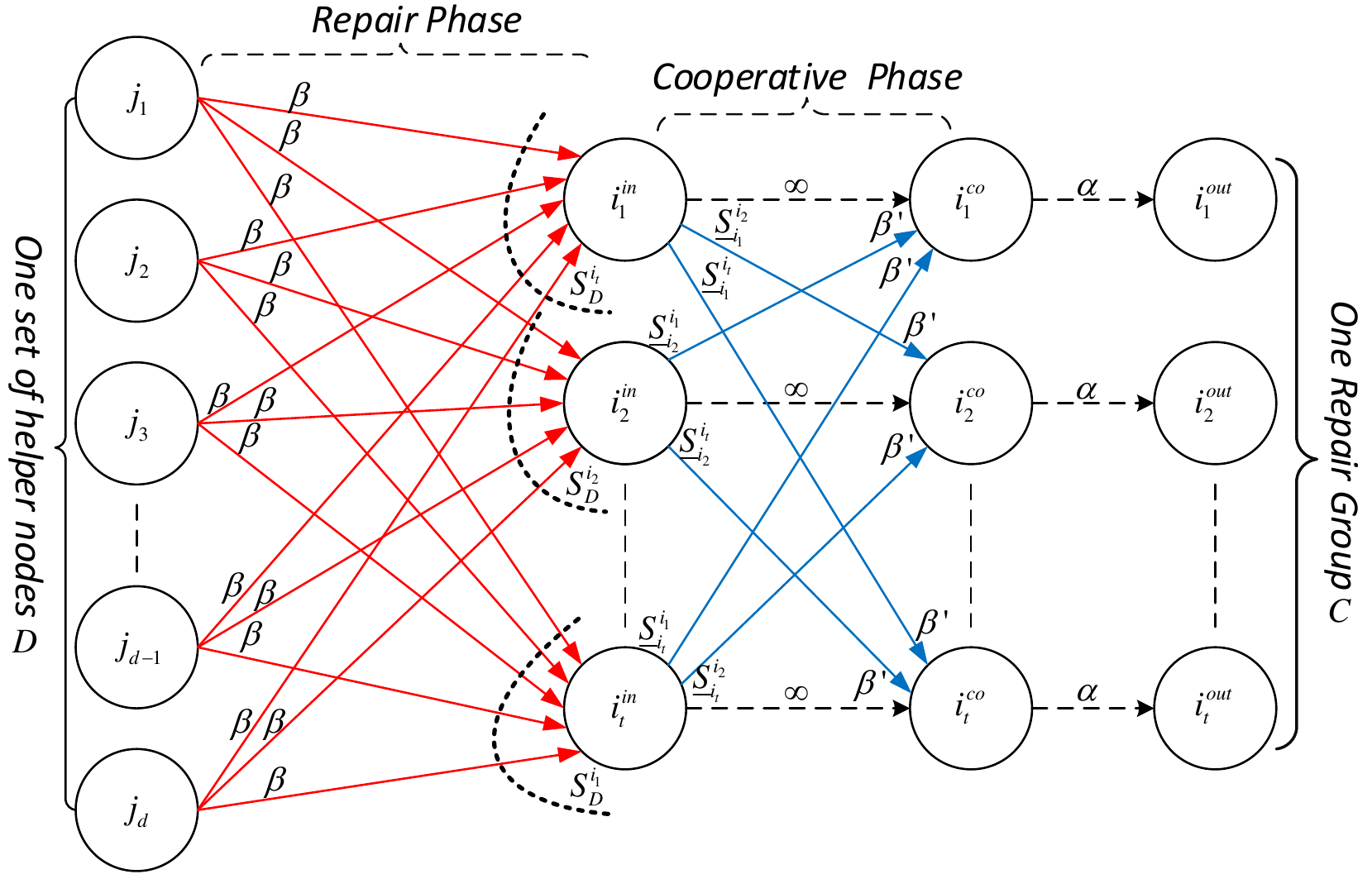}\\
  \caption{$C=(i_1,\cdots,i_t)$ is one repair group and $D=(j_1,j_2,\cdots,j_d)$ is one set of helper nodes, where $C$ is disjoint with $D$. In the first repair phase, each new node in $C$ downloads $\beta$ symbols from each helper node in $D$, i.e., $(S_D^{i_1},\cdots,S_D^{i_t})$. In the cooperative repair phase, each new node mutually exchange $\beta'$ symbols, i.e., $(\underline{S}_{C\setminus \{i_1\}}^{i_1},\cdots,\underline{S}_{C\setminus \{i_t\}}^{i_t})$. Thus, the total repair downloads for each new node in $C$ is $\{S_D^{i_l},\underline{S}_{C\setminus \{i_l\}}^{i_l}\}$ for $1\leq l\leq t$, which is used to recover $W_{i_l}$ the original storage of failed node $i_l$.}\label{Fig:system model}
\end{figure}

The following is the parameter notation of cooperative regenerating codes $\{n\geq d+t,k,d,t,\alpha,\beta,\beta'\}$, which is reduced to the scenario of regenerating codes when $t=1$. Based on the repair process, there are totally ${n \choose t}$ possible different repair groups. Fig.\ref{Fig:system model} describes the basic system model with some parameters.

{\bf B. \emph{Parameter Notations}:} Given any cooperative regenerating code with parameter set $\{n\geq d+t,k,d,t,\alpha,\beta,\beta'\}$, we let

(1) $W_i, i\in[1,n]$ denote the random variable corresponding to the content of node $i$, which has that $H(W_i)=\alpha$.

(2) $\{W_A, A\subseteq [1,n]\}$ denote the set of random variables corresponding to the nodes in the subset $A$. Throughout the paper, subscripts of $W$ can represent either a node index or a set of nodes which will be clear from the context.

(3) $S_i^j,\{i,j\}\subset[1,n],i\neq j$ denote the random variable corresponding to the symbols of \textbf{repair data} sent by the surviving node $i$ to new node $j$, where $H(S_i^j)=\beta$.

(4) $S_A^B$ denote the set $\{S_i^j|i\in A,j\in B,i\neq j,A\subseteq [1,n],B\subseteq [1,n]\}$, and particularly $S^B$ substitutes for $S_{[1,n]}^B$.

(5) $\underline{S}_i^j,\{i,j\}\subset[1,n],i\neq j$ denote the random variable corresponding to the symbols of \textbf{exchanging data} sent by the new node $i$ to another new node $j$, when node $i$ and node $j$ are in the same repair group, where $H(\underline{S}_i^j)=\beta'$.

(6) $\underline{S}_A^B$ denote the set $\{\underline{S}_i^j|i\in A,j\in B,i\neq j,A\subseteq [1,n],B\subseteq [1,n]\}$.

\begin{rem}
Compared to regenerating codes, cooperative regenerating codes have another parameter that is the exchanging data $\underline{S}_i^j$. According to the above notation of the exchanging data $\underline{S}_i^j$ and the procedure of the cooperative repair, it must be that, for any repair group $C$ and any helper nodes set $D$ where $i\in C$ and $D\subseteq [1,n]\setminus C$,

\begin{equation}\label{basic cond}
\left\{\begin{aligned}
&H(\underline{S}_{i}^{C\setminus \{i\}}|S_D^{i})=0\\
&H(W_i|S_D^{i},\underline{S}_{C\setminus \{i\}}^i)=0,
\end{aligned}\right.
\end{equation}
where the first term means that exchanging data $\underline{S}_{i}^{C\setminus \{i\}}$ is the function of the repair data of node $i$ and the second term implies that node $i$ can be regenerated by the repair data $S_D^{i}$ as well as the exchanging data $\underline{S}_{C\setminus \{i\}}^i$.

In addition, for any $\{n\geq d+t,k,d,t,\alpha,\beta,\beta'\}$ MSCR code, it must be an MDS code (reconstruction property) and have the regeneration property that any $t$ failed nodes can be repaired simultaneously. These two basic properties can be expressed as

\begin{equation}\label{basic property}
\left\{\begin{aligned}
&H\big(\{W_{i_j}\}_{j=1}^k\big)=k\alpha\\
&H(W_C|S_{D}^{C})=0,\\
\end{aligned}\right.
\end{equation}
where $C$ and $D$ are defined as equation (\ref{basic cond}). When $n=d+t$, $D$ is unique after the choice of $C$.
\end{rem}

{\bf C. \emph{Eavesdropping Model}:} We consider an $\{l_1,l_2\}$-eavesdropper, which has access to the storage contents of nodes in set $E$ and additionally can observe the repair downloads of nodes in set $F$, where $|E|=l_1$, $|F|=l_2$ and $l_1+l_2<k$. Besides, we set $G$ to be another nodes set of size $(k-l_1-l_2)$, where $G\subseteq[1,n]\backslash (E\cup F)$.

\begin{figure}[!h]
  \centering
  \includegraphics[width=12 cm]{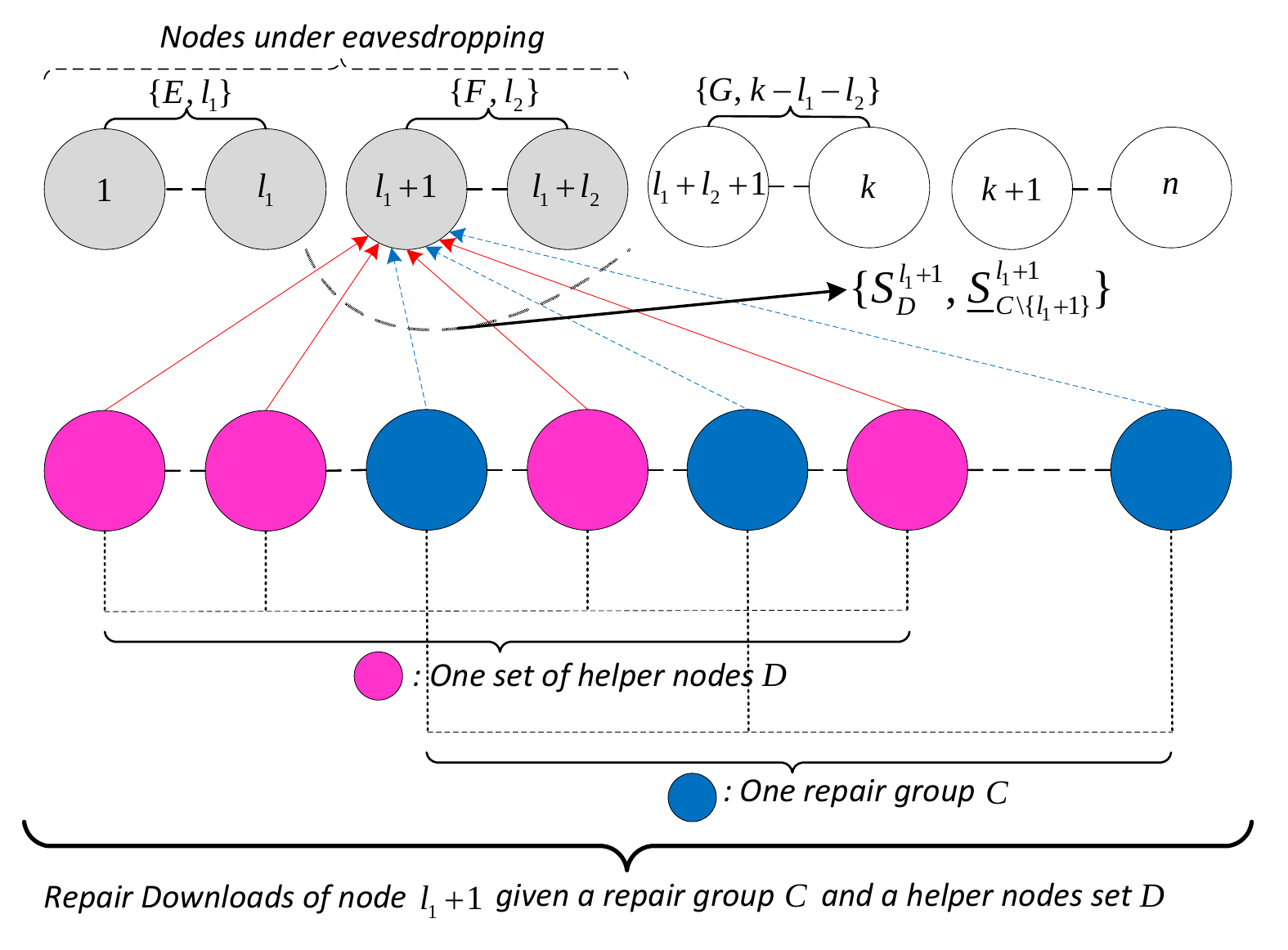}\\
  \caption{$E$ is the nodes set whose contents are eavesdropped and $F$ is the nodes set whose repair downloads are observed by eavesdropper. Given a repair group $C$ including node $l_1+1$ and a set of helper nodes $D$, red lines indicate the repair data $S^{l_1+1}_D$ and blue lines stand for the exchanging data $\underline{S}_{C\setminus \{l_1+1\}}^{l_1+1}$, which constitute the total repair downloads of failed node $l_1+1$. For all possible repair groups and the sets of helper nodes, the repair downloads of node $l_1+1$ that the eavesdropper may obtain is $\{S^{l_1+1}_D,\underline{S}_{C\setminus \{l_1+1\}}^{l_1+1}|l_1+1\in C,C\widetilde{\subset}[1,n],D\widetilde{\subset}[1,n]\setminus C\}$. Thus, for the eavesdropped nodes set $E$ and $F$, the total information may leaked to eavesdropper is $\left\{W_E,\{S_D^l,\underline{S}_{C\setminus \{l\}}^l|l\in F, l\in C, C\widetilde{\subset}[1,n], D\widetilde{\subset}[1,n]\setminus C\}\right\}$. }\label{Fig:eavesdropping}
\end{figure}

However, different from regenerating codes, the repair downloads of any one node in $F$ here are comprised of the repair data from $d$ helper nodes and the exchanging data from $t-1$ cooperative nodes. As shown in Figure. \ref{Fig:eavesdropping}, there are totally ${n-1 \choose t-1}$ possible sets of the cooperative nodes after deciding one failed node and ${n-t \choose d}$ possible helper nodes sets after determining a repair group. Thus, after traversing all possible repair groups and the sets of helper nodes, the $\{l_1,l_2\}$-eavesdropper is supposed to have the knowledge

\begin{equation}
\left\{W_E,\{S_D^i,\underline{S}_{C\setminus \{i\}}^i| i\in C\cap F, C\widetilde{\subset}[1,n], D\widetilde{\subset}[1,n]\setminus C, |C|=t, |D|=d\}\right\},
\end{equation}
where $C\widetilde{\subset}[1,n]$ indicates that $C$ traverses $[1,n]$ and so does $D$. For brevity, we substitute $\{S_D^i,\underline{S}_{C\setminus \{i\}}^i| i\in C\cap F, C\widetilde{\subset}[1,n], D\widetilde{\subset}[1,n]\setminus C, |C|=t, |D|=d\}$ for $\tilde{S}^F$ and thus $\{W_E,\tilde{S}^F\}$ is the data information leakage obtained by eavesdropper. In \cite{Re:Koyluoglu}, the authors only consider the eavesdropping model under single repair group.

\vspace{0.4cm}

{\bf D. \emph{Security Consideration}:} Based on the above eavesdropping model, we consider a special class of MSCR codes, where the repair data sent from any surviving node $i$ to a new node $j$ is independent of the choice of the other $t-1$ cooperative nodes and the other $d-1$ helper nodes. That is to say, the content of single repair data $S_i^j$ is fixed and only depends on the helper node index $i$ and the failed node index $j$. However, we do not restrict the content of exchanging data $\underline{S}_i^j$ also to be invariant, i.e., it may vary depending on different repair groups including both nodes $i$ and $j$. Nevertheless, we will show that it does not matter if the exchanging data is restricted to be fixed or not, when considering the total amount of information leakage.

As discussed before, this restriction of repair data is important for the MSCR codes to be secure, since the $\{l_1,l_2\}$-eavesdropper can get access to the repair downloads of the nodes in $F$. Otherwise, the changing contents of repair data $\{S_i^j, j\in F\}$ will cause more information leakage due to different repair groups or different sets of helper nodes, which is certain analogous to the situation of functional repair and may make it impossible to maintain the security of MSCR codes. Based on this security concern, we define such an MSCR code as

\begin{definition}(Stable MSCR Code): A stable MSCR code with $\{n\geq d+t,k,d,t,\alpha,\beta,\beta'\}$ is an MSCR code with the ``stable" repair property, that is, for arbitrary repair group $C$ including $j$ and arbitrary set of helper nodes $D$ including $i$, the content of repair data $S_i^j$ is independent of the choices of $C$ and $D$, where $i\neq j\in [1,n]$.
\end{definition}

In next section, we will reconsider the two MSCR codes \cite{Re:N. Le Scouarnec,Re:K. W. Shum}, while the authors in \cite{Re:Koyluoglu} only considered under single repair group and neglected this ``stable" property of MSCR codes.

\section{ILLUSTRATION OF EXISTING MSCR CODES}
In this section, we reanalyze the secrecy capacity of the two MSCR codes \cite{Re:N. Le Scouarnec,Re:K. W. Shum}, whose detail on the stable property is overlooked in \cite{Re:Koyluoglu}.  Both MSCR codes \cite{Re:N. Le Scouarnec,Re:K. W. Shum} will be shown not stable. The MSCR code proposed in \cite{Re:N. Le Scouarnec} will be further shown impossible to be transformed into a secure MSCR code under the $\{l_1=0,l_2=1\}$-eavesdropping model. As for the one in \cite{Re:K. W. Shum}, its original repair procedure is also not stable, but it can be converted to a stable one through adjusting the repair strategy.
\subsection{Unstable MSCR Codes}
Here, we take the two MSCR codes as examples and explain why they are not stable and why it is hard or even impossible for them to maintain the data secrecy under the $\{l_1,l_2\}$-eavesdropping model.
\subsubsection{3.1.1  MSCR-Code-A.}

The authors in \cite{Re:Koyluoglu} first investigated the secrecy capacity of the MSCR code \cite{Re:N. Le Scouarnec} with special parameter $\{d\geq k=t=2\}$. Under the constraint that $l_1+l_2<k=2$, they analyzed two cases respectively, i.e., $\{l_1=1,l_2=0\}$ and $\{l_1=0,l_2=1\}$. The first case $\{l_1=1,l_2=0\}$ is trivial, as there is only some node's content undergoing eavesdropped and does not involve the information leakage of repair downloads. Thus, the construction of secure MSCR code under the $\{l_1=1,l_2=0\}$-eavesdropping model given in \cite{Re:Koyluoglu} is correct.

As for the second case $\{l_1=0,l_2=1\}$, they considered under single repair group made of two systematic nodes. However, they overlooked the fact that the content of the repair data transferred for one systematic node, changes with different repair groups which could include the same systematic node but another parity node. In the following, we first describe the coding scheme and the repair strategy as given in \cite{Re:N. Le Scouarnec}, then we show that this code in \cite{Re:N. Le Scouarnec} cannot be transformed into a secure MSCR code under the $\{l_1=0,l_2=1\}$-eavesdropping model.

\vspace{0.2cm}
$\bullet$ {\bf Coding Scheme:} The coding scheme is specified by $\{k=t=2,\beta=1\}$, from which it has the special parameter setting with $\{\alpha=d-k+t=d=n-2, B=k(d-k+t)=2\alpha\}$. Keeping the notation used in \cite{Re:Koyluoglu}, the procedure is described as follows:

$\ast$: $\mathbf{a}=(a_1,a_2,\cdots,a_{\alpha})^T$ is systematically stored in the first node.

$\ast$: $\mathbf{b}=(b_1,b_2,\cdots,b_{\alpha})^T$ is systematically stored in the second node.

$\ast$: $\mathbf{r}_i=(a_1+\omega^{(i-1)\mod \alpha}b_1,\cdots, a_{\alpha}+\omega^{(i+\alpha-2)\mod \alpha}b_{\alpha})^T$ is stored in $i$th parity node, where $i\in [1,d]$ and $\omega$ is the generator of a finite field $\mathbb{F}_q$. For convenient index, the $i$th parity node is marked as the $(i+2)$th node, $i\in [1,d]$. By matrix representation, $\mathbf{r}_i=\mathbf{a}+\mathbf{B}_i\mathbf{b}$, where $\mathbf{B_i}$ is the corresponding diagonal matrix.

$\bullet$ {\bf Repair Strategy:} The detailed coding construction can be referred to \cite{Re:N. Le Scouarnec} and we only care about its repair process. As described in \cite{Re:N. Le Scouarnec}, they only consider the repair group comprised of two systematic nodes. Other repair groups including parity node can be performed as the two systematic nodes after change of variables. Assume the repair downloads of the first node (node $1$) is observed by the $\{l_1=0,l_2=1\}$-eavesdropper. Under repair group $(1,2)$, the repair data sent from the $j$th parity node to node $1$ is given by
\begin{equation}\label{example}
\left\{S_{j+2}^{1_{(1,2)}}:\mathbf{v}_{1,j}^T\mathbf{r}_j=\mathbf{z}^T\mathbf{B}^{-1}_{j}\mathbf{r}_j=\mathbf{z}^T\mathbf{B}^{-1}_{j}\mathbf{a}+\overbrace{\mathbf{z}^T\mathbf{b}}\right\},
\end{equation}
where they set $\mathbf{z}=(1,\cdots,1)^T$ and $\overbrace{\mathbf{z}^T\mathbf{b}}$ is termed an interference needing canceling out.

Now, we consider other situations when a repair group is comprised of the first node and the $i$th parity node where $i\neq j$. As suggested, we should view $\{\mathbf{a},\mathbf{r}_i\}$ or $\{1,i+2\}$ as two systematic nodes. For simplicity, we let $\mathbf{x}=\mathbf{a}$ and $\mathbf{y}=\mathbf{r}_i$. After changing variables, we have
\begin{equation}
\left\{\begin{aligned}
&\mathbf{b}=-\mathbf{B}^{-1}_i\mathbf{x}+\mathbf{B}^{-1}_i\mathbf{y}\\
&\mathbf{r}_j=(\mathbf{I}-\mathbf{B}_{j}\mathbf{B}^{-1}_{i})\mathbf{x}+\mathbf{B}_{j}\mathbf{B}^{-1}_{i}\mathbf{y},
\end{aligned}\right.
\end{equation}
where $\mathbf{I}$ is the identical matrix. In order to ensure the alignment of interference, the $j$th parity node now should send to node $1$ under repair group $(1,i+2)$ by
\begin{equation}
\left\{S_{j+2}^{1_{(1,i+2)}}:\mathbf{v'}^T_{1,j}\mathbf{r}_j=\mathbf{z}^T\mathbf{B}_i\mathbf{B}^{-1}_{j}\mathbf{r}_j
=\mathbf{z}^T(\mathbf{B}_i\mathbf{B}^{-1}_{j}-\mathbf{I})\mathbf{x}+\overbrace{\mathbf{z}^T\mathbf{y}}
=\mathbf{z}^T\mathbf{B}_i(\mathbf{B}^{-1}_{j}\mathbf{a}+\mathbf{b})\right\},
\end{equation}
where $\overbrace{\mathbf{z}^T\mathbf{y}}$ now is viewed as an interference. Similarly, the second systematic node (whose storage is $\mathbf{b}$) should send to node $1$ under repair group $(1,i+2)$ by
\begin{equation}
\left\{S_{2}^{1_{(1,i+2)}}:\mathbf{z}^T\mathbf{B}_i\mathbf{b}
=\mathbf{z}^T\mathbf{B}_i(-\mathbf{B}^{-1}_i\mathbf{x}+\mathbf{B}^{-1}_i\mathbf{y})
=-\mathbf{z}^T\mathbf{x}+\overbrace{\mathbf{z}^T\mathbf{y}}\right\},
\end{equation}
where $\overbrace{\mathbf{z}^T\mathbf{y}}$ needs to be canceled out.

\begin{figure}[!h]
  \centering
  \includegraphics[width=12 cm]{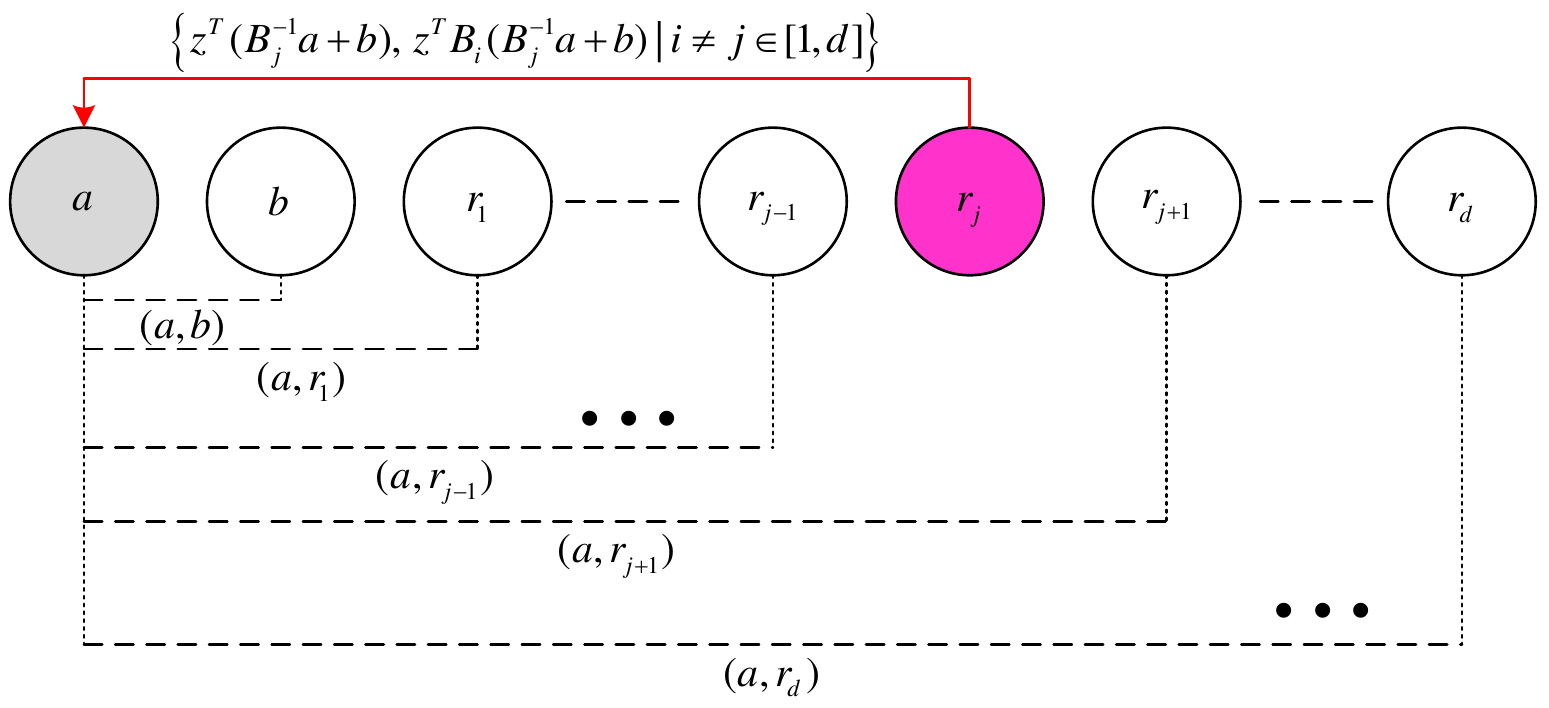}\\
  \caption{Under different repair groups including node $1$, the node $j+2$ (or the $j$th parity node) sends different contents of repair data $S_{j+2}^1$ to node $1$, which will leak more data information to the eavesdropper.}\label{Fig:unstable model}
\end{figure}

$\bullet$ {\bf Data Eavesdropped:} Under the $\{l_1=0,l_2=1\}$-eavesdropping model, when the repair downloads of node $1$ is eavesdropped, the total data eavesdropped (that in fact is all the repair downloads of node $1$ under all possible repair groups) is comprised of the repair data of node $1$ from the helper nodes and the exchanging data from the corresponding cooperative nodes. As shown in \cite{Re:Koyluoglu}, under the single repair group made of two systematic nodes $(1,2)$, the information symbols observed by the eavesdropper is given by
\begin{equation}\label{leak}
\left\{\mathbf{z}^T\big{(}\nu(\omega^0+\omega^{1}\cdots +\omega^{\alpha-1})^{-1}\mathbf{a}+\mathbf{b}\big{)},\mathbf{z}^T(\mathbf{B}^{-1}_1\mathbf{a}+\mathbf{b}),
\mathbf{z}^T(\mathbf{B}^{-1}_2\mathbf{a}+\mathbf{b}),\cdots,\mathbf{z}^T(\mathbf{B}^{-1}_d\mathbf{a}+\mathbf{b})
\right\},
\end{equation}
where $\mathbf{z}^T\big{(}\nu(\omega^0+\omega^{1}\cdots +\omega^{\alpha-1})^{-1}\mathbf{a}+\mathbf{b}\big{)}$ is the exchanging data from node $2$.  Next, we will show that, the already obtained content of node $1$ combined with the repair data sent from any one helper node to node $1$ after traversing all possible repair groups are enough for eavesdropper to retrieve all the data information. In other words, only needing the repair downloads of node $1$ under any one repair group and all the repair data sent from node $j+2$ to node $1$ under all possible repair groups, the eavesdropper can recover all the original data information.

\vspace{0.2cm}
As illustrated in Fig. \ref{Fig:unstable model}, after traversing all possible repair groups including node $1$, the eavesdropper can totally obtain $(d=\alpha)$-sized repair data from the $j$th parity node to the first node that are
\begin{equation}
\left\{[S_{j+2}^{1_{(1,2)}},S_{j+2}^{1_{(1,i+2)}}]=[\mathbf{z}^T(\mathbf{B}^{-1}_{j}\mathbf{a}+\mathbf{b}),\mathbf{z}^T\mathbf{B}_i(\mathbf{B}^{-1}_{j}\mathbf{a}+\mathbf{b})]\mid i\neq j\in [1,d]\right\},
\end{equation}
which is equivalent to
\begin{equation}\label{leakage1}
(\mathbf{a}^T\mathbf{B}^{-1}_{j}+\mathbf{b}^T)\cdot[\mathbf{z},\mathbf{B}_1\mathbf{z},\cdots,\mathbf{B}_{j-1}\mathbf{z},\mathbf{B}_{j+1}\mathbf{z},\cdots,\mathbf{B}_{d}\mathbf{z}].
\end{equation}
Here, it should be noted that the eavesdropper not merely can obtain the repair data sent from the $j$th parity node as the formula (\ref{leakage1}), but also can observe other repair downloads including the repair data and the exchanging data which are sent from other helper nodes and cooperative nodes. Although the information leakage as in the formula (\ref{leakage1}) differs from the formula (\ref{leak}) given in \cite{Re:Koyluoglu}, it is now clear for us that both information leakage formulas (\ref{leak}) and (\ref{leakage1}) actually are only parts of the total data eavesdropped under all possible repair groups. The reason of here only considering the repair data sent from the $j$th parity node as the formula (\ref{leakage1}) is that, the eavesdropper has been able to sufficiently decode the original data information, only using the already known content of $\mathbf{a}$ and these information symbols of the repair data as the formula (\ref{leakage1}). It is illustrated as follows.

\vspace{0.2cm}
Required by the coding construction in \cite{Re:N. Le Scouarnec}, the following $\alpha \times \alpha$ matrix
\begin{equation}\label{matrix1}
[\mathbf{z},\mathbf{B}^{-1}_1\mathbf{z},\cdots,\mathbf{B}^{-1}_{j-1}\mathbf{z},\mathbf{B}^{-1}_{j+1}\mathbf{z},\cdots,\mathbf{B}^{-1}_{d}\mathbf{z}],
\end{equation}
should be invertible, which, as stated in \cite{Re:Koyluoglu}, can be guaranteed by the condition that $q> n-1$ and $(\omega^0+\cdots +\omega^{\alpha-1})^2\omega^{-(\alpha-1)}\notin \{0,\alpha^2\}$. Actually, based on this condition, we can also deduce that the following matrix from the formula (\ref{leakage1})
\begin{equation}\label{matrix2}
[\mathbf{z},\mathbf{B}_1\mathbf{z},\cdots,\mathbf{B}_{j-1}\mathbf{z},\mathbf{B}_{j+1}\mathbf{z},\cdots,\mathbf{B}_{d}\mathbf{z}]
\end{equation}
is invertible\footnote{\emph{Proof}: First, $B_i^{-1}$ is a diagonal matrix whose diagonal elements are $\{\omega^{(1-i)\mod \alpha},\cdots,\omega^{(2-i-\alpha)\mod \alpha}\}$. Then, matrix (\ref{matrix2}) can be equivalently transformed into matrix (\ref{matrix1}), if $\omega^{-1}$ is regarded as the generator of the finite field $\mathbb{F}_q$. At last, if $\omega^{-1}$ satisfies $(\omega^0+\omega^{-1}\cdots +\omega^{1-\alpha})^2\omega^{(\alpha-1)}\notin \{0,\alpha^2\}$, matrix (\ref{matrix2}) is invertible. For this, we can easily find the clue that
$(\omega^0+\omega^{-1}\cdots +\omega^{1-\alpha})^2\omega^{(\alpha-1)}=(\omega^0+\omega^{-1}\cdots +\omega^{1-\alpha})^2(\omega^{(\alpha-1)})^2\omega^{-(\alpha-1)}=
[(\omega^0+\omega^{-1}\cdots +\omega^{1-\alpha})\omega^{(\alpha-1)}]^2\omega^{-(\alpha-1)}=(\omega^0+\cdots +\omega^{\alpha-1})^2\omega^{-(\alpha-1)}$.
}.

Therefore, the eavesdropper can obtain the content of $(\mathbf{a}^T\mathbf{B}^{-1}_{j}+\mathbf{b}^T)$ just only by solving the equation (\ref{leakage1}). In fact, the content of $(\mathbf{a}^T\mathbf{B}^{-1}_{j}+\mathbf{b}^T)$ include all the storage information of node $j+2$, since $\mathbf{r}_j=\mathbf{a}+\mathbf{B}_j\mathbf{b}$. Then, combining the already obtained content of $\mathbf{a}$ under any one repair group, he thus can obtain the content of $\mathbf{b}$. That is to say, the $\{l_1=0,l_2=1\}$-eavesdropper can obtain all the information of the original data message $(\mathbf{a},\mathbf{b})$, as long as by observing the repair downloads of node $1$ which undergoes all the repair groups $(1,l)$ for $l\in [1,d+2]\setminus \{1,j+2\}$. In this case, we cannot implement one-time pad scheme to encrypt or randomize secure information symbols as used in \cite{Re:Koyluoglu}, since all the information symbols have been eavesdropped and there are no secure information symbols left. Hence, this MSCR code in \cite{Re:N. Le Scouarnec} cannot be transformed into a secure MSCR code under the $\{l_1=0,l_2=1\}$-eavesdropping model.

\subsubsection{3.1.2 MSCR-Code-B.}

The authors in \cite{Re:Koyluoglu} then investigated the secrecy capacity of MSCR code given in \cite{Re:K. W. Shum} with $\{d=k, \alpha=t,\beta=1\}$, which actually is also not stable.

$\bullet$ {\bf Coding Deployment:} As shown in \cite{Re:K. W. Shum}, the $k\cdot t$ original data packets are deployed in a $t\times k$ data matrix $\mathbf{M}$ and its row representation is denoted by $(\mathbf{m}^T_1,\mathbf{m}^T_2,\cdots,\mathbf{m}^T_t)$. Consider a $k\times n$ generator matrix
\begin{equation}
\mathbf{G}=\left[
  \begin{array}{ccccc}
    1         & 1         & 1        &\cdots  & 1        \\
    a_1       & a_2       &a_3       &\cdots  &a_n       \\
    \vdots    &\vdots     &\vdots    &\ddots  &\vdots    \\
    a^{k-1}_1 &a^{k-1}_2  &a^{k-1}_3 &\cdots  &a^{k-1}_n \\
  \end{array}
\right],
\end{equation}
of which every $k\times k$ submatrix is a non-singular Vandermonde matrix. Then encode the original data matrix into $\mathbf{MG}$ and the encoded data packets stored in node $j$ are $\{\mathbf{m}^T_i \mathbf{g}_j|i=1,2,\cdots,t\}$, where $\mathbf{g}_j$ is the $j$th column of $\mathbf{G}$.

$\bullet$ {\bf Repair Strategy:} When $t$ nodes are failed, $t$ new nodes contact any other $d=k$ surviving nodes, where the $t$ new nodes are indexed by $\{f_1,\cdots,f_t\}$ and the $k$ helper nodes are indexed by $\{\lambda_1,\cdots,\lambda_k\}$. Each helper node $\lambda_l$ sends its $j$th packet to the new node $f_j$ with $\mathbf{m}^T_j \mathbf{g}_{\lambda_l}$, for $l\in [1,k]$. Because of the property of Vandermonde matrix, $\mathbf{m}^T_j$ can be recovered by reversing the matrix $[\mathbf{g}_{\lambda_1},\mathbf{g}_{\lambda_2},\cdots,\mathbf{g}_{\lambda_k}]$. In the cooperative repair phase, the new node $f_j$ sends $\mathbf{m}^T_j \mathbf{g}_{f_i}$ to another new node $f_i$, for $i\neq j\in [1,t]$. Thus, the new node $f_j$ can receive $t-1$ data packets $\{\mathbf{m}^T_i \mathbf{g}_{f_j}|i\neq j\in [1,t]\}$ during cooperative repair phase. Combining the previously obtained $\mathbf{m}^T_j$, the initial state of node $f_j$ can be recovered.

\begin{figure}[!h]
  \centering
  \includegraphics[width=10 cm]{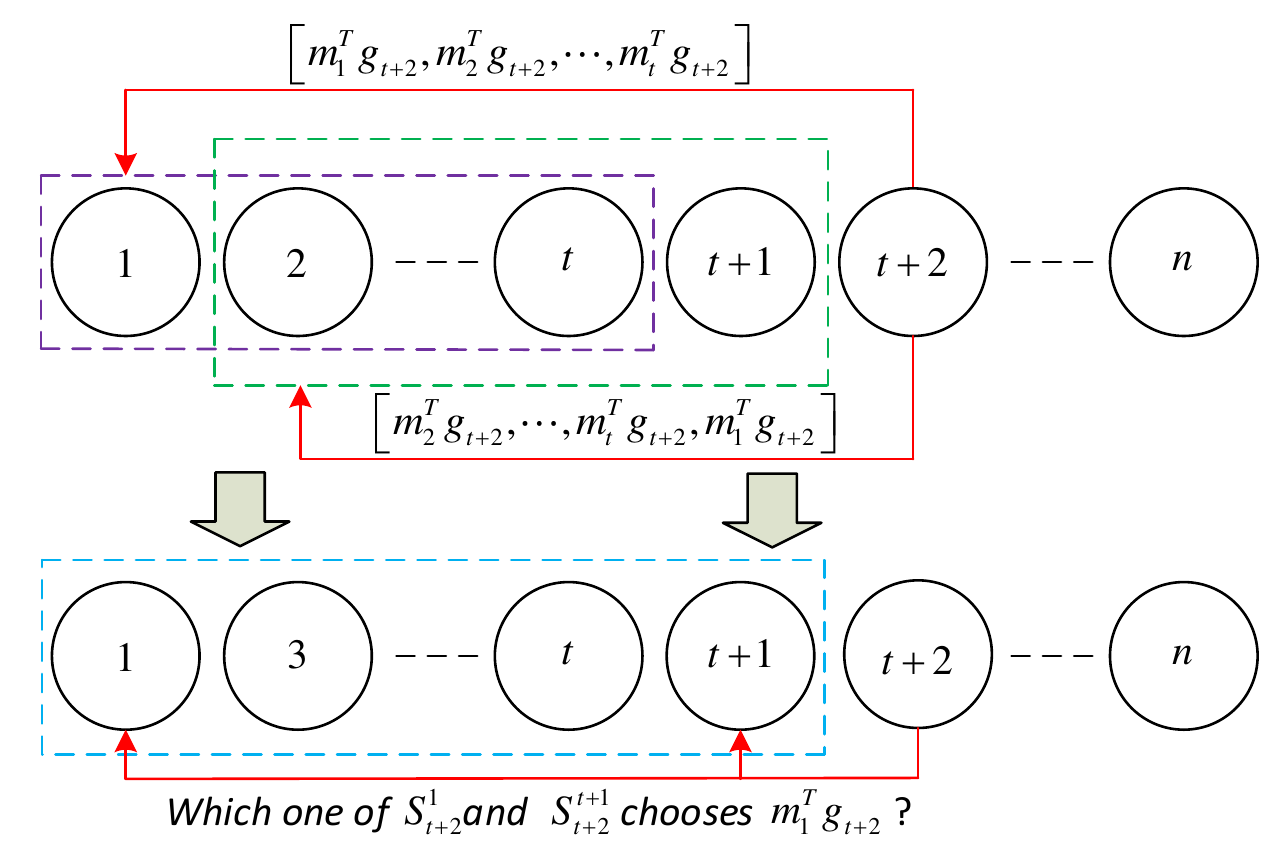}\\
  \caption{In the repair group $[1,t]$, $S_{t+2}^1=\mathbf{m}^T_1 \mathbf{g}_{t+2}$. In the repair group $[2,t+1]$, $S_{t+2}^{t+1}=\mathbf{m}^T_1 \mathbf{g}_{t+2}$. However, in the repair group $[1,3,\cdots,t+1]$, we can only set $\{S_{t+2}^1=\mathbf{m}^T_1 \mathbf{g}_{t+2}, S_{t+2}^{t+1}=\mathbf{m}^T_2 \mathbf{g}_{t+2}\}$ or $\{S_{t+2}^1=\mathbf{m}^T_2 \mathbf{g}_{t+2}, S_{t+2}^{t+1}=\mathbf{m}^T_1 \mathbf{g}_{t+2}\}$, which indicates one of repair data $S_{t+2}^1$ and $S_{t+2}^{t+1}$ must change its content. If the eavesdropper observes repair downloads of the node that has changing contents of repair data, it will obviously obtain more data information.}\label{Fig:unstable1 model}
\end{figure}

$\bullet$ {\bf Data Eavesdropped:} According to the repair process, we find that the repair data from a helper node $\lambda_l$ to a new node $f_j$ is $\mathbf{m}^T_j \mathbf{g}_{\lambda_l}$, where $j\in [1,t]$ and $f_j\in[1,n]$. That implies the mapping of $f_j$ is not bijective. Besides, there are totally ${n\choose t}$ possible repair groups. So, there must exist two different repair groups $\{f_1,\cdots,f_t\}$ and $\{f'_1,\cdots,f'_t\}$, where $f_j\neq f'_j$ and $\{S_{\lambda_l}^{f_j}=S_{\lambda_l}^{f'_j}=\mathbf{m}^T_j \mathbf{g}_{\lambda_l}\}$ for some $j$. However, when node $f_j$ and $f'_j$ are in the same repair group, $S_{\lambda_l}^{f_j}$ and $S_{\lambda_l}^{f'_j}$ cannot be equal to $\mathbf{m}^T_j \mathbf{g}_{\lambda_l}$ simultaneously. In other words, we cannot guarantee that repair data from any helper node to any failed node is always fixed, which exactly means this MSCR code is not stable and will leak more data information if the eavesdropper can observe the repair downloads of the corresponding node.

\vspace{0.2cm}
As shown in Fig. \ref{Fig:unstable1 model}, for repair group $[1,t]$, we set $S_{t+2}^{i_{[1,t]}}=\mathbf{m}^T_i \mathbf{g}_{t+2}$ for $i\in [1,t]$. For another repair group $[2,t+1]$, we set $S_{t+2}^{{t+1}_{[2,t+1]}}=\mathbf{m}^T_1 \mathbf{g}_{t+2}$ and $\{S_{t+2}^{i_{[2,t+1]}}=\mathbf{m}^T_i \mathbf{g}_{t+2}$ for $i\in [2,t]$. However, when node $1$ and node $t+1$ are in the same repair group such as $[1,3,\cdots,t+1]$, $S_{t+2}^{1_{[1,3,\cdots,t+1]}}$ and $S_{t+2}^{{t+1}_{[1,3,\cdots,t+1]}}$ cannot equal with $\mathbf{m}^T_1 \mathbf{g}_{t+2}$ simultaneously.

\vspace{0.4cm}

As stated in \cite{Re:K. W. Shum}, any $t$ new nodes are put in order by their serial numbers. In fact, such an order arrangement is the least secure way. For example, if $n\geq 2t+k-1$, when repair group $[1,t]$ gradually traverse to repair group $[t,2t-1]$, the repair data sent to node $t$ from helper nodes set $[2t,2t+k-1]$ is given by
\begin{equation}
\{S_{\lambda}^{{t}_{[i,t-1+i]}}=\mathbf{m}^T_{t+1-i} \mathbf{g}_{\lambda}\mid i\in [1,t],\lambda\in[2t,2t+k-1]\}=\{\mathbf{m}^T_{1} \mathbf{g}_{\lambda},\cdots,\mathbf{m}^T_{t} \mathbf{g}_{\lambda}|\lambda\in[2t,2t+k-1]\},
\end{equation}
which, if observed by eavesdropper, can be used to decode all the original data packets $(\mathbf{m}_1,\mathbf{m}_2,\cdots,\mathbf{m}_t)$ since $[\mathbf{g}_{2t},\cdots,\mathbf{g}_{2t+k-1}]$ is invertible. It means that the eavesdropper can obtain all the original data information only by observing the repair data of node $t$ involved in repair groups as many as possible.

\begin{rem}
Although the MSCR code given in \cite{Re:K. W. Shum} is not stable and possesses poor secrecy capacity, it can be converted to a stable one by adjusting its repair strategy, which will offer better secrecy capacity.
\end{rem}

\subsection{A Stable MSCR Code}
In this section, we will present a stable MSCR code built from conversion of repair strategy based on the MSCR code given in \cite{Re:K. W. Shum}.
\vspace{0.2cm}

We apply the same coding deployment but change the repair strategy, where the main purpose is to make the content of repair data $S_{\lambda_l}^{f_j}$ invariant to the choice of helper node $\lambda_l$ and failed node $f_j$. In other words, we need to ensure the bijection between indices of failed nodes and repair data packets given by a helper node. Thus, after the coding deployment, we consider a systematic MDS code $(\mathbf{m}'_1,\mathbf{m}'_2,\cdots,\mathbf{m}'_t,\mathbf{m}'_{t+1},\cdots,\mathbf{m}'_n)$ which is extended by the original data packets $(\mathbf{m}_1,\mathbf{m}_2,\cdots,\mathbf{m}_t)$, where $(\mathbf{m}'_1,\mathbf{m}'_2,\cdots,\mathbf{m}'_t)=(\mathbf{m}_1,\mathbf{m}_2,\cdots,\mathbf{m}_t)$. For this, we can use a $t\times n$ generator matrix $\mathbf{G}'$
\begin{equation}
\mathbf{G'}=\left[
  \begin{array}{cccccccc}
    1         & 0       &\cdots           & 0        &\nu_{1,t+1}   &\nu_{1,t+2}   &\cdots  & \nu_{1,n}    \\
    0         & 1       &\cdots           &0          &\nu_{2,t+1}  &\nu_{2,t+2}   &\cdots   & \nu_{2,n}   \\
    \vdots    &\vdots   &\ddots           &\vdots    &\vdots        &\vdots        &\ddots    & \vdots     \\
    0          &0       & \cdots         &1           &\nu_{t,t+1}  &\nu_{t,t+2}   &\cdots    & \nu_{t,n}  \\
  \end{array}
\right],
\end{equation}
of which every $t\times t$ submatrix is invertible. We let $\mathbf{g}'_j$ denotes the $j$th column of $\mathbf{G'}$. Here, it should be noted that $\mathbf{G}$ is a $k\times n$ matrix, while $\mathbf{G'}$ is a $t\times n$ matrix. So, we have
\begin{equation}\label{conver}
[\mathbf{m}_1,\mathbf{m}_2,\cdots,\mathbf{m}_t] \cdot \mathbf{G}'=[\mathbf{m}'_1,\mathbf{m}'_2,\cdots,\mathbf{m}'_t,\mathbf{m}'_{t+1},\cdots,\mathbf{m}'_n],
\end{equation}
from which we can derive, for any $i\in[1,n-t]$,
\begin{equation}
\mathbf{m}'_{t+i}=[\mathbf{m}_1,\mathbf{m}_2,\cdots,\mathbf{m}_t] \cdot \mathbf{g}'_{t+i}=\nu_{1,t+i}\mathbf{m}_1+\nu_{2,t+i}\mathbf{m}_2+\cdots+\nu_{t,t+i}\mathbf{m}_t.
\end{equation}

\begin{figure}[!h]
  \centering
  \includegraphics[width=12 cm]{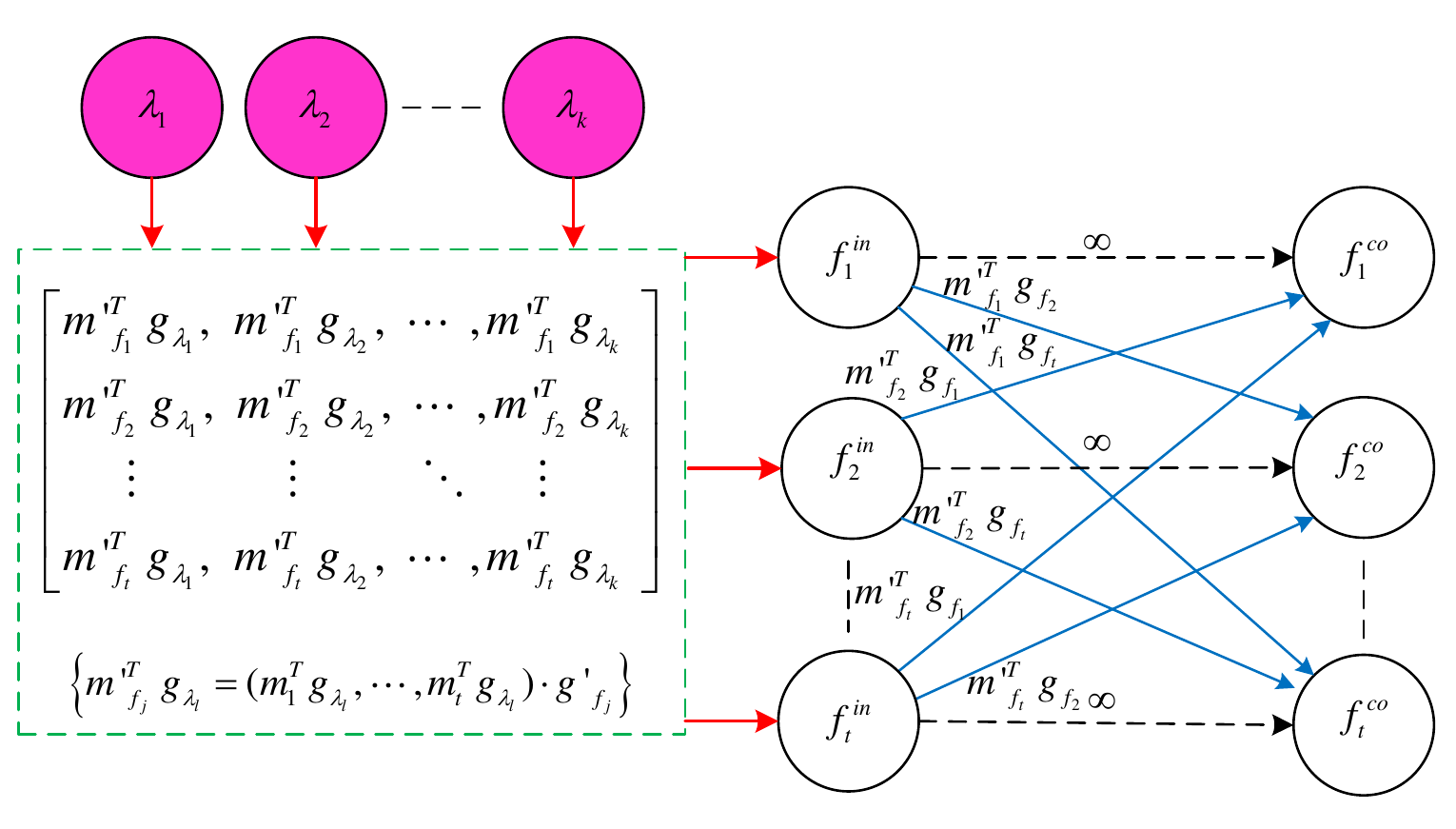}\\
  \caption{Given a repair group $\{f_1,\cdots,f_t\}$ and a set of helper nodes $\{\lambda_1,\cdots,\lambda_k\}$, $\mathbf{m}'^T_{f_j} \mathbf{g}_{\lambda_l}$ is the repair data sent from node $\lambda_l$ to node $f_j$. Subsequently, each new node $f_j$ sends $\mathbf{m}'^T_{f_j} \mathbf{g}_{f_i}$ to another new node $f_i$, where $i\neq j\in [1,t]$. Each new node is then recovered exactly, by combining all the repair data and the exchanging data.}\label{Fig:stable model}
\end{figure}

The following is the new repair strategy which is also shown in Fig. \ref{Fig:stable model}.

\vspace{0.2cm}

\emph{\textbf{Step 1}}. For any repair group $\{f_1,\cdots,f_t\}$ and any set of helper nodes $\{\lambda_1,\cdots,\lambda_k\}$, each helper node $\lambda_l$ sends to the new node $f_j$ with $(\underline{\mathbf{m}^T_1\mathbf{g}_{\lambda_l}, \cdots,\mathbf{m}^T_t\mathbf{g}_{\lambda_l}})\cdot \mathbf{g}'_{f_j}$, where
\begin{equation}\label{conversion}
\left\{\begin{aligned}
&(\underline{\mathbf{m}^T_1\mathbf{g}_{\lambda_l}, \cdots,\mathbf{m}^T_t\mathbf{g}_{\lambda_l}})\cdot \mathbf{g}'_{f_j}\\
&=\mathbf{g}'^T_{f_j} \cdot (\mathbf{m}^T_1\mathbf{g}_{\lambda_l}, \cdots,\mathbf{m}^T_t\mathbf{g}_{\lambda_l})^T\\
&=\mathbf{g}'^T_{f_j} \cdot [\mathbf{m}_1,\mathbf{m}_2,\cdots,\mathbf{m}_t]^T \cdot \mathbf{g}_{\lambda_l}\\
&=([\mathbf{m}_1,\mathbf{m}_2,\cdots,\mathbf{m}_t]\cdot \mathbf{g}'_{f_j})^T \cdot \mathbf{g}_{\lambda_l}\\
&=\mathbf{m}'^T_{f_j} \mathbf{g}_{\lambda_l},
\end{aligned}\right.
\end{equation}
where $(\underline{\mathbf{m}^T_1\mathbf{g}_{\lambda_l}, \cdots,\mathbf{m}^T_t\mathbf{g}_{\lambda_l}})$ is the exact original storage of node $\lambda_l$ and $\mathbf{m}'^T_{f_j}$ is from equation (\ref{conver}). So, the repair data $\{S_{\lambda_l}^{f_j}=\mathbf{m}'^T_{f_j} \mathbf{g}_{\lambda_l}\}$ now actually is the linear combination of storage in node $\lambda_l$, while the original repair data is $\mathbf{m}^T_{j} \mathbf{g}_{\lambda_l}$ (the $j$th data packet of node $\lambda_l$). Furthermore, due to the invertiblity of any $k\times k$ submatrix $[\mathbf{g}_{\lambda_1},\cdots,\mathbf{g}_{\lambda_k}]$ of $\mathbf{G}$, the linear combination of original data $\mathbf{m}'^T_{f_j}$ is obtained.
\vspace{0.2cm}

\emph{\textbf{Step 2}}. In the cooperative repair phase, the new node $f_j$ sends exchanging data $\mathbf{m}'^T_{f_j} \mathbf{g}_{f_i}$ to other new nodes $f_i$, for $i\neq j\in [1,t]$. Hence, the new node $f_j$ can receive $t-1$ data packets $\{\mathbf{m}'^T_{f_i} \mathbf{g}_{f_j}|i\neq j\in [1,t]\}$ in this phase.
\vspace{0.2cm}

\emph{\textbf{Step 3}}. At last, node $f_j$ combines the repair data and exchanging data $\{\mathbf{m}'^T_{f_j},\mathbf{m}'^T_{f_i} \mathbf{g}_{f_j}|i\neq j\in [1,t]\}$ to obtain $\{\mathbf{m}'^T_{f_i} \mathbf{g}_{f_j}|i\in [1,t]\}$, which can be further expressed as
\begin{equation}
\left\{\begin{aligned}
&[\mathbf{m}'_{f_1},\mathbf{m}'_{f_2},\cdots,\mathbf{m}'_{f_t}]^T \cdot  \mathbf{g}_{f_j}\\
&=\left\{[\mathbf{m}_1,\mathbf{m}_2,\cdots,\mathbf{m}_t]\cdot [ \mathbf{g}'_{f_1},\mathbf{g}'_{f_2},\cdots,\mathbf{g}'_{f_t}]\right\}^T \cdot  \mathbf{g}_{f_j}\\
&=[ \mathbf{g}'_{f_1},\mathbf{g}'_{f_2},\cdots,\mathbf{g}'_{f_t}]^T \cdot [\mathbf{m}_1,\mathbf{m}_2,\cdots,\mathbf{m}_t]^T \cdot  \mathbf{g}_{f_j}\\
&=[ \mathbf{g}'_{f_1},\mathbf{g}'_{f_2},\cdots,\mathbf{g}'_{f_t}]^T \cdot (\underline{\mathbf{m}^T_1\mathbf{g}_{f_j}, \cdots,\mathbf{m}^T_t\mathbf{g}_{f_j}})^T,
\end{aligned}\right.
\end{equation}
where $(\underline{\mathbf{m}^T_1\mathbf{g}_{f_j}, \cdots,\mathbf{m}^T_t\mathbf{g}_{f_j}})$ is the original storage of node $f_j$. As any $t\times t$ submatrix $[ \mathbf{g}'_{f_1},\mathbf{g}'_{f_2},\cdots,\mathbf{g}'_{f_t}]$ of $\mathbf{G}'$ is invertible, node $f_j$ can be recovered.

\begin{rem}
According to the above new repair strategy, it is obvious that the content of repair data from any helper node $\lambda_l$ to any failed node $f_j$ $(S_{\lambda_l}^{f_j}=\mathbf{m}'^T_{f_j} \mathbf{g}_{\lambda_l}=(\underline{\mathbf{m}^T_1\mathbf{g}_{\lambda_l}, \cdots,\mathbf{m}^T_t\mathbf{g}_{\lambda_l}})\cdot \mathbf{g}'_{f_j})$ is independent of repair groups and sets of helper nodes. So, this MSCR code built from conversion of repair strategy is a stable MSCR code.
\end{rem}

In subsequent discussion, we study the secrecy capacity of stable MSCR codes from information theoretic perspective. Besides, we will use the above stable MSCR code to calculate its specific secrecy capacity.

\section{INFORMATION THEORETIC FEATURES OF MSCR CODES}
In this section, we first present a generally applicable secrecy expression for MSCR codes. Then, we present some information theoretic features based on the basic reconstruction and regeneration properties of general MSCR and stable MSCR codes.

\subsection{Expression of Secrecy Capacity}
 As assumed in eavesdropping model, the $\{l_1,l_2\}$-eavesdropper has access to the following information
\begin{equation}
\{W_E,\tilde{S}^F\}=\left\{W_E,\{S_D^i,\underline{S}_{C\setminus \{i\}}^i| i\in C\cap F, C\widetilde{\subset}[1,n], D\widetilde{\subset}[1,n]\setminus C, |C|=t, |D|=d\}\right\}.
\end{equation}
Similar to the definition of secrecy capacity of MSR codes \cite{Kun}, we have the following result.
\begin{lemma}\label{secure expression}
For any MSCR code with parameter set $\{n\geq d+t,k,d,t,\alpha,\beta,\beta'\}$, we have
\begin{equation}
\left\{\begin{aligned}
&B^{(s)}\\
&= H(W_E,W_F,W_G|W_E,\tilde{S}^F)\\
&=H(W_G|W_E,W_F)-H(\tilde{S}^F|W_E,W_F)\\
&=(k-l_1-l_2)\alpha-H(\tilde{S}^F|W_E,W_F)
\end{aligned}\right.
\end{equation}
\end{lemma}

\begin{proof}

First, we can use the MRD codes \cite{Re:R. M. Roth} (e.g. Gabidulin code \cite{Re:Gabidulin}) to pre-code the original data file of size $\{B=k\alpha\}$, which is required to consist of $\{B-H(W_E,\tilde{S}^F)\}$-sized secure data file and $H(W_E,\tilde{S}^F)$-sized random data file. As shown in \cite{Re:N.B.Shah,Re:Rawat.A.S,Re:Koyluoglu}, this kind of construction of secure codes always can meet the \emph{conditions of secrecy}\footnote{Consider a DSS with
data file $\mathbf{f}^s$, random data file $\mathbf{r}$ (independent of $\mathbf{f}^s$), and an eavesdropper with observations given by $\mathbf{e}$. If $H(\mathbf{e})\leq H(\mathbf{r})$ and $H(\mathbf{r}|\mathbf{f}^s,\mathbf{e})=0$, then the mutual information leakage to eavesdropper is zero, i.e., $I(\mathbf{f}^s;\mathbf{e})=0$.}, which exactly means the maximal file size that can be securely stored is
\begin{equation}
B^{(s)}=B-H(W_E,\tilde{S}^F)= H(W_E,W_F,W_G|W_E,\tilde{S}^F).
\end{equation}

Second, we can deduce
\begin{equation}
\left\{\begin{aligned}
&H(W_G|W_E,W_F)-H(W_E,W_F,W_G|W_E,\tilde{S}^F)\\
&=H(W_G|W_E,W_F)-H(W_E,W_F,W_G|W_E,W_F,\tilde{S}^F)\\
&=H(W_G|W_E,W_F)-H(W_G|W_E,W_F,\tilde{S}^F)\\
&=I(W_G;\tilde{S}^F|W_E,W_F)\\
&=H(\tilde{S}^F|W_E,W_F)-H(\tilde{S}^F|W_E,W_F,W_G)\\
&=H(\tilde{S}^F|W_E,W_F).
\end{aligned}\right.
\end{equation}

Then, for the MSCR codes, we further have $H(W_G|W_E,W_F)=(k-l_1-l_2)\alpha$, where $\alpha=(d-k+t)\beta$.
Combining these equations, we get the proof.
\end{proof}

\begin{rem}
Based on this definition of secrecy capacity, we only need to calculate or estimate the value of $H(\tilde{S}^F|W_E,W_F)$.
\end{rem}

\subsection{Properties of General MSCR Codes}

We present some properties of MSCR codes as below.

\begin{lemma}\label{conditional entropy}
For any MSCR code with parameter set $\{n\geq d+t,k,d,t,\alpha,\beta,\beta'\}$ where $t\leq k$, consider any three pairwise disjoint subsets $A$, $B$ and $C$ with $\{|C|=t,|A|=k-t,|B|=d-k+t\}$, it must be that
\begin{equation}\label{MSCR pro}
\left\{ \begin{array}{l}
H(S_{A\cup B}^C)=dt\beta\\
H(S_{B}^C|W_C,S_{A}^C)=0.
\end{array} \right.
\end{equation}
\end{lemma}

\begin{proof} We present them as follows.

1. Because MSCR codes are the storage efficient codes with the MDS property, it is trivial that $H(W_C|S_A^C)=H(W_C)=t\alpha$ since $|A|+|C|=k$ and $A\cap C={\O}$.

\vspace{0.2cm}
2. Set $B=\{b_1,b_2,\cdots, b_{d-k+t}\}$. From equation (\ref{basic property}), we know $H(W_C|S_{A\cup B}^C)=0$. Now, we have
\begin{equation}\label{inequation}
\left\{ \begin{aligned}
&H(W_C|S_{A}^C)-H(W_C|S_{A}^C,S_{b_1}^C)\\
&=I(W_C;S_{b_1}^C|S_{A}^C)\\
&=H(S_{b_1}^C|S_{A}^C)-H(S_{b_1}^C|W_C,S_{A}^C)\\
&\leq H(S_{b_1}^C)\\
&\leq t\beta;\\
&\quad \quad\vdots\quad\quad\quad\quad\quad\quad\quad\vdots\\
&H(W_C|S_{A}^C,S_{b_1}^C,S_{b_2}^C,\cdots,S_{b_{d-k+t-1}}^C)-H(W_C|S_{A}^C,S_{B}^C)\\
&=I(W_C;S_{b_{d-k+t}}^C|S_{A}^C,S_{B\setminus\{{b_{d-k+t}}\}}^C)\\
&=H(S_{b_{d-k+t}}^C|S_{A}^C,S_{B\setminus\{{b_{d-k+t}}\}}^C)-H(S_{b_{d-k+t}}^C|W_C,S_{A}^C,S_{B\setminus\{{b_{d-k+t}}\}}^C)\\
&\leq H(S_{b_{d-k+t}}^C)\\
&\leq t\beta.
\end{aligned} \right.
\end{equation}
By summing up the inequalities, we derive
\begin{equation}
t\alpha=H(W_C|S_{A}^C)-H(W_C|S_{A}^C,S_{B}^C)\leq (d-k+t)t\beta.
\end{equation}

Because $\alpha=(d-k+t)\beta$, it is mandatory that all the inequalities (\ref{inequation}) actually are equations. Thus, for $1\leq i\leq d-k+t$, we all have
\begin{equation}\label{addition pro1}
\left\{ \begin{array}{l}
H(S_{b_i}^C|S_{A}^C,S_{\{b_1,\cdots,b_{i-1}\}}^C)=t\beta\\
H(S_{b_i}^C|W_C,S_{A}^C,S_{\{b_1,\cdots,b_{i-1}\}}^C)=0,
\end{array} \right.
\end{equation}
from which we further obtain
\begin{equation}\label{condi}
\left\{\begin{aligned}
&H(S_{B}^C|S_{A}^C)\\
&=\sum_{i=1}^{i=d-k+t}H(S_{b_i}^C|S_{A}^C,S_{\{b_1,\cdots,b_{i-1}\}}^C)\\
&=(d-k+t)t\beta\\
\end{aligned}\right.
\end{equation}
and
\begin{equation}\label{condtional 1}
\left\{\begin{aligned}
&H(S_{B}^C|W_C,S_{A}^C)\\
&=\sum_{i=1}^{i=d-k+t}H(S_{b_i}^C|W_C,S_{A}^C,S_{\{b_1,\cdots,b_{i-1}\}}^C)\\
&=0.
\end{aligned}\right.
\end{equation}

According to equation (\ref{condi}), we further know $H(S_{B}^C)=(d-k+t)t\beta$, with which we obtain $H(S_{b}^C)=t\beta$ for any $b\in B$. Due to the randomness of the choice of the two sets $A$ and $B$, we can also deduce $H(S_A^C)=(k-t)t\beta$ for $|A|=k-t<k$. Thus, combining equation (\ref{condi}), we get
\begin{equation}\label{union entropy}
\left\{\begin{aligned}
&H(S_{A\cup B}^C)\\
&=H(S_{A}^C)+H(S_{B}^C|S_{A}^C)\\
&=(d-k+t)t\beta+(k-t)t\beta\\
&=dt\beta.
\end{aligned}\right.
\end{equation}

Based on the above proof, it is obvious that equation (\ref{MSCR pro}) still holds, when $t=k$ and $A={\O}$.

\end{proof}

\begin{rem}\label{general class}
Since it is trivial that $H(S_{A\cup B}^C)\leq dt\beta$, equation (\ref{union entropy}) exactly means that there are no intersection pattern within the repair data $S_{A\cup B}^C$, i.e., all the contents of repair data $S^C_{A\cup B}$ are mutually independent when $t\leq k$. In addition, we have the following observations:

1. When $t\leq k$,  equation (\ref{union entropy}) further implies that $dt\beta\leq k\alpha$ as the total information entropy of data storage is $k\alpha$, which leads to $(d-k)(k-t)\beta\geq 0$. When $k>t$, it must be that $d\geq k$. When $t=k$, if $d<k$, the two terms of equation (\ref{basic property}) will be contradictory. Thus, it must be that $d\geq k$ when $t\leq k$.

2. When $t>k$, the second term of equation (\ref{basic property}) $H(W_C|S_D^C)=0$ means that $k\alpha\leq dt\beta$, which is equivalent to $(d-k)(t-k)\beta\geq0$. Hence, it also can be derived that $d\geq k$ in this case.

3. Both cases show that there do not exist MSCR codes with $d<k$.

Furthermore, it is interesting to find that when $t\geq k$ and $d=k$, it must be that $H(S_D^C)=dt\beta$, because $\alpha=(d-k+t)\beta=t\beta$ which leads to $k\alpha=H(W_C)\leq H(S_D^C)\leq dt\beta=k\alpha$. In other words, there are also no intersection pattern within the repair data $S_D^C$ when $t\geq k$ and $d=k$.
\end{rem}

\begin{lemma}\label{single entropy}
For any MSCR code with parameter set $\{n\geq d+t,k,d,t,\alpha,\beta,\beta'\}$, consider any single repair of node $i$ in a repair group $\{i,C'\}$ and two other disjoint subsets $A'$ and $B'$ such that $\{|C'|=t-1,|A'|=k-1,|B'|=d-k+1, (A'\cup B')\cap C'={\O}, i\notin \{A'\cup B'\cup C'\}\}$, it must be that
\begin{equation}\label{MSCR pro1}
\left\{ \begin{array}{l}
H(S_{A'\cup B'}^i,\underline{S}_{C'}^i)=(d+t-1)\beta\\
H(S_{B'}^i,\underline{S}_{C'}^i|W_i,S_{A'}^i)=0.
\end{array} \right.
\end{equation}
\end{lemma}

\begin{proof}
We let $B'=\{b'_1,\cdots,b'_{d-k+1}\}$ and $C'=\{c'_1,\cdots,c'_{t-1}\}$. Then, we have

\begin{equation}\label{inequation1}
\left\{ \begin{aligned}
&H(W_i|S_{A'}^i)-H(W_i|S_{A'}^i,S_{b'_1}^i)\\
&=I(W_i;S_{b'_1}^i|S_{A'}^i)\\
&=H(S_{b'_1}^i|S_{A'}^i)-H(S_{b'_1}^i|W_i,S_{A'}^i)\\
&\leq H(S_{b'_1}^i)\\
&\leq \beta;\\
&\quad \quad\vdots\quad\quad\quad\quad\quad\quad\quad\vdots\\
&H(W_i|S_{A'}^i,S_{b'_1}^i,S_{b'_2}^i,\cdots,S_{b'_{d-k}}^i)-H(W_i|S_{A'}^i,S_{B'}^i)\\
&=I(W_i;S_{b'_{d-k+1}}^i|S_{A'}^i,S_{B'\setminus\{{b'_{d-k+1}}\}}^i)\\
&=H(S_{b'_{d-k+1}}^i|S_{A'}^i,S_{B'\setminus\{{b'_{d-k+1}}\}}^i)-H(S_{b'_{d-k+1}}^i|W_i,S_{A'}^i,S_{B'\setminus\{{b'_{d-k+1}}\}}^i)\\
&\leq H(S_{b'_{d-k+1}}^i)\\
&\leq \beta;
\end{aligned} \right.
\end{equation}
and
\begin{equation}\label{inequation2}
\left\{ \begin{aligned}
&H(W_i|S_{A'}^i,S_{B'}^i)-H(W_i|S_{A'}^i,S_{B'}^i,\underline{S}_{c'_1}^i)\\
&=I(W_i;\underline{S}_{c'_1}^i|S_{A'\cup B'}^i)\\
&=H(\underline{S}_{c'_1}^i|S_{A'\cup B'}^i)-H(\underline{S}_{c'_1}^i|W_i,S_{A'\cup B'}^i)\\
&\leq H(\underline{S}_{c'_1}^i)\\
&\leq \beta';\\
&\quad \quad\vdots\quad\quad\quad\quad\quad\quad\quad\vdots\\
&H(W_i|S_{A'\cup B'}^i,\underline{S}_{c'_1}^i,\cdots,\underline{S}_{c'_{t-2}}^i)-H(W_i|S_{A'\cup B'}^i,\underline{S}_{C'}^i)\\
&=I(W_i;\underline{S}_{c'_{t-1}}^i|S_{A'\cup B'}^i,\underline{S}_{C'\setminus \{c'_{t-1}\}}^i)\\
&=H(\underline{S}_{c'_{t-1}}^i|S_{A'\cup B'}^i,\underline{S}_{C'\setminus \{c'_{t-1}\}}^i)-H(\underline{S}_{c'_{t-1}}^i|W_i,S_{A'\cup B'}^i,\underline{S}_{C'\setminus \{c'_{t-1}\}}^i)\\
&\leq H(\underline{S}_{c'_{t-1}}^i)\\
&\leq \beta'.
\end{aligned} \right.
\end{equation}

By summing up all the inequalities (\ref{inequation1}) and (\ref{inequation2}) along with the fact that $\beta=\beta'$ in the MSCR scenario, we derive
\begin{equation}
\alpha=H(W_i|S_{A'}^i)-H(W_i|S_{A'\cup B'}^i,\underline{S}_{C'}^i)\leq (d+t-k)\beta,
\end{equation}
from which all the inequalities (\ref{inequation1}) and (\ref{inequation2}) mandatorily become the equations similar to Lemma \ref{conditional entropy}. Thus, we get the proof.
\end{proof}

\begin{rem}
According to the second term of equation (\ref{MSCR pro1}), we naturally derive
\begin{equation}
\left\{\begin{aligned}\label{equi}
&H(S_{B'}^i|W_i,S_{A'}^i)=0\\
&H(\underline{S}_{C'}^i|W_i,S_{A'}^i)=0,
\end{aligned}\right.
\end{equation}
using which we can further simplify $H(\tilde{S}^F|W_E,W_F)$.

\end{rem}

\subsection{Properties of Stable MSCR Codes}
Some properties of stable MSCR codes are present as follows. Here, we should know that stable MSCR codes also have the above properties of general MSCR codes in Lemma \ref{conditional entropy} and \ref{single entropy}, since stable MSCR codes still are MSCR codes.

\begin{lemma}\label{secure}
For any stable MSCR code with parameter set $\{n\geq d+t,k,d,t,\alpha,\beta,\beta'\}$, we have\footnote{Lemma \ref{secure} shows that it does not matter if the exchanging data in the stable MSCR scenario is restricted to be fixed or not, because the exchanging data $\{\underline{S}_{C\setminus \{i\}}^i| i\in C, C\widetilde{\subset}[1,n]\}$ is only a function of the content of $\{W_i,S_{[1,n]\setminus \{i\}}^i\}$. In other words, the total information of the exchanging data $\{\underline{S}_{C\setminus \{i\}}^i| i\in C, C\widetilde{\subset}[1,n]\}$ is included in $\{W_i,S_{[1,n]\setminus \{i\}}^i\}$. So, when calculating the amount of the eavesdropped data, we do not need to consider the exchanging data $\{\underline{S}_{C\setminus \{i\}}^i| i\in C, C\widetilde{\subset}[1,n]\}$, while we only need to focus on the combination of some node's storage and its repair data $\{W_i,S^i\}$.}
\begin{equation}
\tilde{S}^F=\{W_F,S^F\},
\end{equation}
from which we further obtain
\begin{equation}
H(\tilde{S}^F|W_E,W_F)=H(S^F|W_E,W_F)=H(S_G^F|W_E,W_F)=H(S_G^F),
\end{equation}
where $G$ is a set of size $(k-l_1-l_2)$ and is disjoint with $E$ and $F$ as defined in the eavesdropping model.
\end{lemma}

\begin{proof}
The proof is separated into two parts as below.
\vspace{0.2cm}

\textbf{1}. First, we know for any $i\in F$, $\tilde{S}^i=\{S_D^i,\underline{S}_{C\setminus \{i\}}^i| i\in C, C\widetilde{\subset}[1,n], D\widetilde{\subset}[1,n]\setminus C, |C|=t, |D|=d\}$. The ``stable" property of MSCR codes will lead to that
\begin{equation}
\tilde{S}^i=\{S_{[1,n]\setminus \{i\}}^i,\underline{S}_{C\setminus \{i\}}^i| i\in C, C\widetilde{\subset}[1,n],|C|=t\},
\end{equation}
where we claim again that exchanging data $\{\underline{S}_{j}^i|i,j\in C\}$ does not have the ``stable" constraints and may vary depending on different repair groups $C$. In addition, it must be that $H(W_i,S^i|\tilde{S}^i)=0$ from equation (\ref{basic cond}). The following shows that the exchanging data $\{\underline{S}_{C\setminus \{i\}}^i| C\widetilde{\subset}[1,n],|C|=t\}$ is a function of the content of $\{W_i,S_{[1,n]\setminus \{i\}}^i\}$, where $S_{[1,n]\setminus \{i\}}^i$ can be replaced by $S^i$.

For any repair group $C$ including $i$, there always exists some set $A''$ such that $A''\cap C={\O}$ and $|A''|=k-1$, because $d\geq k$. Then, according to the second term of equation (\ref{equi}) in Lemma \ref{single entropy}, we have
\begin{equation}
H(\underline{S}_{C\setminus \{i\}}^i|W_i,S_{A''}^i)=0.
\end{equation}
Thereby, we derive
\begin{equation}\label{equation entropy}
\left\{ \begin{aligned}
&H(\tilde{S}^i|W_i,S^i)\\
&=H(\underline{S}_{\left\{C\setminus \{i\}| i\in C, C\widetilde{\subset}[1,n]\right\}}^i|W_i,S^i)\\
&=H(\underline{S}_{\left\{C\setminus \{i\}| i\in C, C\widetilde{\subset}[1,n]\right\}}^i|W_i,S_{A''}^i,S_{[1,n]\setminus \{i\cup A''\}}^i)\\
&=0.
\end{aligned} \right.
\end{equation}

Therefore, from $H(W_i,S^i|\tilde{S}^i)=H(\tilde{S}^i|W_i,S^i)=0$, we naturally have $\{\tilde{S}^i\}=\{W_i,S^i\}$ and further get $\{\tilde{S}^F\}=\{W_F,S^F\}$.
\vspace{0.2cm}

\textbf{2}. Assume all the $n$ nodes are comprised of $E,F,G,T$, where $|E\cup F\cup G|=k$ and $|T|=n-k$. So, we have
\begin{equation}
\left\{\begin{aligned}
&H(\tilde{S}^F|W_{\{E,F\}})\\
&=H(W_F,S^F|W_{\{E,F\}})\\
&=H(S^F|W_{\{E,F\}})\\
&=H(S_{E,F,G,T}^F|W_{\{E,F\}})\\
&=H(S_{G,T}^F|W_{\{E,F\}})\\
&=H(S_G^F|W_{\{E,F\}})+H(S_T^F|W_{\{E,F\}},S_G^F).
\end{aligned}\right.
\end{equation}

Then for any $i\in F$,
\begin{equation}
\left\{\begin{aligned}
&H(S_T^i|W_{\{E,F\}},S_G^F)\\
&\leq H(S_T^i|W_{\{E,F\}},S_G^i)\\
&=H(S_T^i|W_i,W_{\{E,F\}\setminus \{i\}},S_G^i)\\
&\leq H(S_T^i|W_i,S_{\{E,F\}\setminus \{i\}}^i,S_G^i)\\
&=H(S_T^i|W_i,S_{\{E,F,G\}\setminus \{i\}}^i).
\end{aligned}\right.
\end{equation}
Based on the first term of equation (\ref{equi}) and the fact that $|\{E,F,G\}\setminus \{i\}|=k-1$, we obtain
\begin{equation}
H(S_{T'}^i|W_{\{E,F\}},S_G^F)=0,
\end{equation}
where $T'$ can be any subset of $T$ of size $d-k+1$. Owing to the randomness of $T'$, we can deduce that $H(S_{T}^i|W_{\{E,F\}},S_G^F)=0$, which further leads to $H(S_T^F|W_{\{E,F\}},S_G^F)=0$. Furthermore, it is trivial that $H(S_G^F|W_E,W_F)=H(S_G^F)$.

\end{proof}

\begin{rem}\label{note1}
From the above proof, we can easily find that the formulation $H(\tilde{S}^F|W_F)=H(S_G^F)$ still holds, when $E={\O}$ and $|F\cup G|=k$. However, it should be noted that, unlike MSR codes, MSCR codes do not necessarily have the property that $H(W_i|S^i)=0$. MSCR codes only have a similar format that $H(W_i|\tilde{S}^i)=0$ instead.
\end{rem}

\begin{lemma}\label{interesting1}
In the stable MSCR scenario, for any subset $F$ such that $|F|\leq k-1$, and arbitrary different $i_1,i_2$ where $i_1,i_2\notin F$, we have $H(S_{i_1}^F)=H(S_{i_2}^F)$. Furthermore, we have

$\bullet$ When $t\leq k$, for any $|F|\leq t$, we always have $H(S_{i}^F)=|F|\beta$, where $i\notin F$.

$\bullet$ When $t\geq k$ and $d=k$,\footnote{In the situation when $t\geq k$ and $d=k$, we should know that if $k\leq|F|\leq t$, the formulation that $H(S_{i}^F)=|F|\beta$ still holds.} for any $|F|\leq t$, we still have $H(S_{i}^F)=|F|\beta$, where $i\notin F$.
\end{lemma}

\begin{proof}
We present them as the following two parts.
\vspace{0.2cm}

\textbf{1}. From Lemma \ref{secure} and Remark \ref{note1}, we have
\begin{equation}\label{general eq}
\left\{\begin{aligned}
&H(\tilde{S}^F)\\
&=H(S^F,W_F)\\
&=H(W_F)+H(S^F|W_F)\\
&=H(W_F)+H(S_{G'}^F|W_F)\\
&=H(W_F)+H(S_{G'}^F),
\end{aligned}\right.
\end{equation}
where $G'$ is a random subset of $[1,n]$ such that $|G'\cup F|=k$ and $G' \cap F={\O}$.
Since $|F|\leq k-1$, then $|G'|\geq 1$.

When $|G'|=1$, for any two different $g_1$ and $g_2$ where $g_1,g_2\in \{[1,n]\setminus F\}$,
\begin{equation}
H(\tilde{S}^F)=H(W_F)+H(S_{g_1}^F)=H(W_F)+H(S_{g_2}^F),
\end{equation}
which indicates $H(S_{g_1}^F)=H(S_{g_2}^F)$.

When $|G'|\geq 2$, we set $G'=\{g',G_1\}$ and $G''=\{g'',G_1\}$ such that $\{g'\neq g'',|G'|=|G''|=k-|F|,G'\cap F=G''\cap F={\O}\}$. Similarly, we obtain
\begin{equation}\label{reduction}
\left\{\begin{aligned}
&H(\tilde{S}^F)\\
&=H(W_F)+H(S_{G'}^F)\\
&=H(W_F)+H(S_{g'}^F)+H(S_{G_1}^F);\\
&H(\tilde{S}^F)\\
&=H(W_F)+H(S_{G''}^F)\\
&=H(W_F)+H(S_{g''}^F)+H(S_{G_1}^F),
\end{aligned}\right.
\end{equation}
which implies $H(S_{g'}^F)=H(S_{g''}^F)$.

Because of the randomness of choices of $(g_1,g_2)$ and $(g',g'')$, we have $H(S_{i_1}^F)=H(S_{i_2}^F)$ for arbitrary different $i_1,i_2$ where $i_1,i_2\notin F$.

\vspace{0.2cm}

\textbf{2}. Remark \ref{general class} in Lemma \ref{conditional entropy} shows that in the situations when $t\leq k$ or when $t\geq k$ and $d=k$, contents of any repair data (from any helper nodes set $D$ to any repair group $C$) are mutually independent. Due to the random choices of $C$ and $D$ and the stable repair property, we obtain for any $|F|\leq t$,  $H(S_{i}^F)=|F|\beta$, where $i\notin F$.

\end{proof}

\section{MAIN RESULTS ON SECRECY CAPACITY}

In this section, we will use a simple formulation to present a generally applicable expression of secrecy capacity for stable MSCR codes. Then, we give some specific results on the secrecy capacity of stable MSCR codes. At last, we take the stable MSCR code as an example to verify the secrecy capacity obtained from information theory.

\subsection{Simple Expression of Secrecy Capacity}
Leveraging the lemmas we obtain before, we have the following theorem.

\begin{theorem}\label{Secure size further}
For any stable MSCR code with parameter set $\{n\geq d+t,k,d,t,\alpha,\beta,\beta'\}$,
\begin{equation}\label{secure size further}
B^{(s)}=(k-l_1-l_2)(\alpha-H(S_g^F)),
\end{equation}
where $g\in G$, $|G|=k-l_1-l_2$ and $|F|=l_2\leq l_1+l_2\leq k-1$.
\end{theorem}

\begin{proof}
Lemma \ref{secure expression} and Lemma \ref{secure} mean that, for the stable MSCR codes, we have the following expression of secrecy capacity
\begin{equation}\label{ea1}
B^{(s)}=(k-l_1-l_2)\alpha-H(S_G^F),
\end{equation}
where $|G|=k-l_1-l_2$ and $l_1+l_2\leq k-1$.

Lemma \ref{interesting1} indicates that, in the stable MSCR scenario, for any subset $F$ such that $|F|\leq k-1$ and for arbitrary $g_1,g_2\in G$, we have
\begin{equation}\label{ea2}
H(S_{g_1}^F)=H(S_{g_2}^F).
\end{equation}

From the equations (\ref{ea1}) and (\ref{ea2}), we naturally obtain the expression (\ref{secure size further}).

\end{proof}

\begin{rem}
The formulation (\ref{secure size further}) can be regarded as the simplest way to define the secrecy capacity of stable MSCR codes, since we only need to concentrate on $S_g^F$, the repair data sent from single node $g$, where $g\in G$.
\end{rem}

\subsection{Some Results on Secrecy Capacity}
Putting all together, we give the following result.

\begin{theorem}\label{final expression}
Given a stable MSCR code with $\{n\geq d+t,k,d,t,\alpha,\beta,\beta'\}$, for $l_1+l_2\leq k-1$, we have
\begin{equation}
B^{(s)}=(k-l_1-l_2)(\alpha-\pi(\beta,l_2)),
\end{equation}
where

\begin{equation}
\pi(\beta,l_2)=l_2\beta, \quad\textrm{for}\quad
\left\{\begin{array}{ll}
& l_2\leq t\leq k;\\
& \textrm{or} \quad t>k \quad\textrm{and}\quad d=k.\\
\end{array}\right.
\end{equation}

\end{theorem}

\begin{proof}
Lemma \ref{interesting1} and Theorem \ref{Secure size further} directly lead to
\begin{equation}\label{tight}
B^{(s)}=(k-l_1-l_2)(\alpha-l_2\beta)=(k-l_1-l_2)(d-k+t-l_2)\beta,
\end{equation}
when $l_2\leq t\leq k$ or when $t> k$ and $d=k$.
\end{proof}

\begin{rem}
The above theorem is only applicable to stable MSCR codes. The authors in \cite{Re:Koyluoglu} give a similar result in the situation when $d=k$ and $l_2\leq t$, while they only consider under single repair group.
\end{rem}

\subsection{Specific Calculation of Secrecy Capacity}
Here, we are to analyze the specific secrecy capacity of the stable MSCR code obtained in Section 3.2.
\vspace{0.2cm}

Without loss of generality, we assume the eavesdropper can observe the content of nodes set $\{E=[1,l_1]\}$ and the repair downloads of nodes set $\{F=[l_1+1,l_1+l_2]\}$, where $l_1+l_2\leq k-1$. Thus, the eavesdropper has the knowledge of
\begin{equation}
\left\{W_{[1,l_1]};\tilde{S}^{[l_1+1,l_1+l_2]}=\{S_D^i,\underline{S}_{\{C\setminus i\}}^i| i\in C\cap [l_1+1,l_1+l_2], C\widetilde{\subset}[1,n], D\widetilde{\subset}([1,n]\setminus C)\}\right\},
\end{equation}
where $C$ denotes the repair group, $D$ is the set of helper nodes and $\widetilde{\subset}$ means traversing. Interestingly, we find that $\underline{S}_j^i=\mathbf{m}'^T_j \mathbf{g}_i$ is also invariant in this stable MSCR code, while we assume it may vary with different repair groups. We make the calculation in detail as follows.

First, we have $W_{[1,l_1]}=\{\mathbf{m}^T_i \mathbf{g}_j|i=1,2,\cdots,t;j=1,\cdots,l_1\}$, where $(\mathbf{m}_1,\mathbf{m}_2,\cdots,\mathbf{m}_t)$ is the original data packets.

Second, we have
\begin{equation}
\left\{\begin{aligned}
&\tilde{S}^{[l_1+1,l_1+l_2]}\\
&=S^{[l_1+1,l_1+l_2]}\cup \underline{S}^{[l_1+1,l_1+l_2]}\\
&=\left\{\mathbf{m}'^T_i \cdot [\mathbf{g}_1,\cdots,\mathbf{g}_{i-1},\mathbf{g}_{i+1},\cdots,\mathbf{g}_{n}],[\mathbf{m}'_1,\cdots,\mathbf{m}'_{i-1},\mathbf{m}'_{i+1},\cdots,\mathbf{m}'_{n}]^T\cdot \mathbf{g}_i |i\in [l_1+1,l_1+l_2]\right\}.\\
&=\left\{\mathbf{g}'^T_{i} \cdot [\mathbf{m}_1,\mathbf{m}_2,\cdots,\mathbf{m}_t]^T \cdot [\mathbf{g}_1,\cdots,\mathbf{g}_{i-1},\mathbf{g}_{i+1},\cdots,\mathbf{g}_{n}]|i\in [l_1+1,l_1+l_2]\right\} \\
&\cup \left\{[\mathbf{g}'_1,\cdots,\mathbf{g}'_{i-1},\mathbf{g}'_{i+1},\cdots,\mathbf{g}'_{n}]^T \cdot [\mathbf{m}_1,\mathbf{m}_2,\cdots,\mathbf{m}_t]^T \cdot \mathbf{g}_{i}|i\in [l_1+1,l_1+l_2]  \right\},
\end{aligned}\right.
\end{equation}
where
\begin{equation}
\left\{\begin{aligned}
&S^{[l_1+1,l_1+l_2]}=\left\{\mathbf{g}'^T_{i} \cdot [\mathbf{m}_1,\mathbf{m}_2,\cdots,\mathbf{m}_t]^T \cdot [\mathbf{g}_1,\cdots,\mathbf{g}_{i-1},\mathbf{g}_{i+1},\cdots,\mathbf{g}_{n}]|i\in [l_1+1,l_1+l_2]\right\}\\
&\underline{S}^{[l_1+1,l_1+l_2]}= \left\{[\mathbf{g}'_1,\cdots,\mathbf{g}'_{i-1},\mathbf{g}'_{i+1},\cdots,\mathbf{g}'_{n}]^T \cdot [\mathbf{m}_1,\mathbf{m}_2,\cdots,\mathbf{m}_t]^T \cdot \mathbf{g}_{i}|i\in [l_1+1,l_1+l_2]  \right\}.
\end{aligned}\right.
\end{equation}

Now, we are to verify some properties of stable MSCR codes.
\vspace{0.2cm}

\emph{\textbf{Verification 1}}. According to the first part of Lemma \ref{secure}, we should have
\begin{equation}
\tilde{S}^{[l_1+1,l_1+l_2]}=\{W_{[l_1+1,l_1+l_2]},S^{[l_1+1,l_1+l_2]}\},
\end{equation}
where $W_{[l_1+1,l_1+l_2]}=[\mathbf{m}_1,\mathbf{m}_2,\cdots,\mathbf{m}_t]^T \cdot [\mathbf{g}_{l_1+1},\cdots,\mathbf{g}_{l_1+l_2}]$. Since any  $t\times t$ submatrix of $\mathbf{G'}$ is invertible, we can directly deduce
\begin{equation}
\left\{\begin{aligned}
&H(\underline{S}^{[l_1+1,l_1+l_2]}|W_{[l_1+1,l_1+l_2]})=0\\
&H(W_{[l_1+1,l_1+l_2]}|\underline{S}^{[l_1+1,l_1+l_2]})=0,
\end{aligned}\right.
\end{equation}
that naturally leads to $\underline{S}^{[l_1+1,l_1+l_2]}=W_{[l_1+1,l_1+l_2]}$ and further verifies the first part of Lemma \ref{secure}
\begin{equation}\label{verify}
\tilde{S}^{[l_1+1,l_1+l_2]}=S^{[l_1+1,l_1+l_2]}\cup \underline{S}^{[l_1+1,l_1+l_2]}=\{W_{[l_1+1,l_1+l_2]},S^{[l_1+1,l_1+l_2]}\}.
 \end{equation}
Here, it should be noted that the property $\underline{S}^{[l_1+1,l_1+l_2]}=W_{[l_1+1,l_1+l_2]}$ is not applicable to any stable MSCR codes and is only feasible in this special stable MSCR code\footnote{Although Lemma \ref{secure} leads to $\tilde{S}^{[l_1+1,l_1+l_2]}=S^{[l_1+1,l_1+l_2]}\cup \underline{S}^{[l_1+1,l_1+l_2]}=\{W_{[l_1+1,l_1+l_2]},S^{[l_1+1,l_1+l_2]}\}$ that corresponds to equation (\ref{verify}), we cannot derive that $\underline{S}^{[l_1+1,l_1+l_2]}=W_{[l_1+1,l_1+l_2]}$ for any stable MSCR codes. The reason is that $\underline{S}^{[l_1+1,l_1+l_2]}$ and $W_{[l_1+1,l_1+l_2]}$ are not independent with $S^{[l_1+1,l_1+l_2]}$, i.e., there exists the intersection pattern between $\underline{S}^{[l_1+1,l_1+l_2]}$ and $S^{[l_1+1,l_1+l_2]}$ as well as between $W_{[l_1+1,l_1+l_2]}$ and $S^{[l_1+1,l_1+l_2]}$.
}.
\vspace{0.2cm}

\emph{\textbf{Verification 2}}. Then, we have $\{W_{[1,l_1]},\tilde{S}^{[l_1+1,l_1+l_2]}\}=\{W_{[1,l_1+l_2]},S^{[l_1+1,l_1+l_2]}\}$, which is the information leakage obtained by the eavesdropper. From the second part of Lemma \ref{secure}, it should be that
\begin{equation}
\left\{\begin{aligned}
&H(W_{[1,l_1+l_2]},S^{[l_1+1,l_1+l_2]})\\
&=H(W_{[1,l_1+l_2]})+H(S^{[l_1+1,l_1+l_2]}|W_{[1,l_1+l_2]})\\
&=H(W_{[1,l_1+l_2]})+H(S_{[l_1+l_2+1,k]}^{[l_1+1,l_1+l_2]}).
\end{aligned}\right.
\end{equation}

As we know,
\begin{equation}
\left\{\begin{aligned}
&W_{[1,l_1+l_2]}=[\mathbf{m}_1,\mathbf{m}_2,\cdots,\mathbf{m}_t]^T \cdot [\mathbf{g}_{1},\cdots,\mathbf{g}_{l_1+l_2}]\\
&S^{[l_1+1,l_1+l_2]}=\left\{\mathbf{g}'^T_{i} \cdot [\mathbf{m}_1,\mathbf{m}_2,\cdots,\mathbf{m}_t]^T \cdot [\mathbf{g}_1,\cdots,\mathbf{g}_{i-1},\mathbf{g}_{i+1},\cdots,\mathbf{g}_{n}]|i\in [l_1+1,l_1+l_2]\right\},
\end{aligned}\right.
\end{equation}
from which we have
\begin{equation}
\left\{\begin{aligned}
&H(S^{[l_1+1,l_1+l_2]}|W_{[1,l_1+l_2]})\\
&=H(S_{[l_1+l_2+1,k]}^{[l_1+1,l_1+l_2]}|W_{[1,l_1+l_2]})+H(S_{[k+1,n]}^{[l_1+1,l_1+l_2]}|W_{[1,l_1+l_2]},S_{[l_1+l_2+1,k]}^{[l_1+1,l_1+l_2]}).
\end{aligned}\right.
\end{equation}
Because any $k\times k$ submatrix of $\mathbf{G}$ is invertible, we know that $[\mathbf{g}_{1},\cdots,\mathbf{g}_{l_1+l_2}]$ and $[\mathbf{g}_{l_1+l_2+1},\cdots,\mathbf{g}_{k}]$ are mutually independent. Based on this observation, we can derive $H(S_{[l_1+l_2+1,k]}^{[l_1+1,l_1+l_2]}|W_{[1,l_1+l_2]})=H(S_{[l_1+l_2+1,k]}^{[l_1+1,l_1+l_2]})$. In addition, given the following formulations
\begin{equation}
\left\{\begin{aligned}
&W_{[1,l_1+l_2]}=[\mathbf{m}_1,\mathbf{m}_2,\cdots,\mathbf{m}_t]^T \cdot [\mathbf{g}_{1},\cdots,\mathbf{g}_{l_1+l_2}]\\
&S_{[l_1+l_2+1,k]}^{[l_1+1,l_1+l_2]}=\left\{[\mathbf{g}'_{l_1+1},\cdots,\mathbf{g}'_{l_1+l_2}]^T \cdot [\mathbf{m}_1,\mathbf{m}_2,\cdots,\mathbf{m}_t]^T \cdot [\mathbf{g}_{l_1+l_2+1},\cdots,\mathbf{g}_{k}]\right\},
\end{aligned}\right.
\end{equation}
we can obtain $[\mathbf{g}'_{l_1+1},\cdots,\mathbf{g}'_{l_1+l_2}]^T \cdot [\mathbf{m}_1,\mathbf{m}_2,\cdots,\mathbf{m}_t]^T$ for the invertiblity of $[\mathbf{g}_{1},\cdots,\mathbf{g}_{k}]$, with which we further derive $\left\{S_{[k+1,n]}^{[l_1+1,l_1+l_2]}=[\mathbf{g}'_{l_1+1},\cdots,\mathbf{g}'_{l_1+l_2}]^T \cdot [\mathbf{m}_1,\mathbf{m}_2,\cdots,\mathbf{m}_t]^T \cdot [\mathbf{g}_{k+1},\cdots,\mathbf{g}_{n}]\right\}$. That exactly means  $H(S_{[k+1,n]}^{[l_1+1,l_1+l_2]}|W_{[1,l_1+l_2]},S_{[l_1+l_2+1,k]}^{[l_1+1,l_1+l_2]})=0$. Thus, the second part of Lemma \ref{secure} is also verified.

\vspace{0.2cm}

\emph{\textbf{Verification 3}}. Finally, we can easily deduce that the size of information leakage obtained by the eavesdropper is precisely equal to

\begin{equation}\label{example1}
\left\{\begin{aligned}
&H(W_{[1,l_1]},\tilde{S}^{[l_1+1,l_1+l_2]})\\
&=H(W_{[1,l_1+l_2]})+H(S_{[l_1+l_2+1,k]}^{[l_1+1,l_1+l_2]})\\
&=(l_1+l_2)\alpha+\sum_{g=l_1+l_2+1}^{k} H(S_{g}^{[l_1+1,l_1+l_2]}),
\end{aligned}\right.
\end{equation}
where $S_{g}^{[l_1+1,l_1+l_2]}=[\mathbf{g}'_{l_1+1},\cdots,\mathbf{g}'_{l_1+l_2}]^T \cdot [\mathbf{m}_1,\mathbf{m}_2,\cdots,\mathbf{m}_t]^T \cdot \mathbf{g}_g$.
Because any $t\times t$ submatrix of $\mathbf{G'}$ is invertible, we have
\begin{equation}\label{example11}
H(S_{g}^{[l_1+1,l_1+l_2]})=
\left\{\begin{aligned}
&l_2\beta \quad& \textrm{if} \quad l_2\leq t;\\
&t\beta  \quad & \textrm{if} \quad l_2\geq t.
\end{aligned}\right.
\end{equation}

Combining equations (\ref{example1}) and (\ref{example11}), we obtain, for $l_1+l_2\leq k-1$,
\begin{equation}\label{final re}
B^{(s)}=
\left\{\begin{aligned}
&(k-l_1-l_2)(\alpha-l_2\beta) \quad& \textrm{if} \quad l_2\leq t;\\
&\quad \quad \quad\quad 0  \quad & \textrm{if} \quad l_2\geq t,
\end{aligned}\right.
\end{equation}
where $\alpha=(d-k+t)\beta=t\beta$.

As we can see, this above result is exactly one special case of our Theorem \ref{final expression} when $d=k$.

\begin{rem}
As shown in section 3.1.2, the original MSCR code given in \cite{Re:K. W. Shum} has poor secrecy capacity and may lose all the data secrecy in some cases even when $l_2=1$. In contrast, the stable MSCR code built from conversion apparently offers better secrecy capacity and always provides the positive secrecy capacity whenever $l_2<t$ and $l_1+l_2\leq k-1$, see equation (\ref{final re}).
\end{rem}

\section{CONCLUSION}

In this work, we study the secrecy capacity of minimum storage cooperative regenerating codes. We recognize a critical detail of the repair strategy, that is, the content of repair data may vary depending on the choice of the repair group or the set of helper nodes, which was neglected by the previous studies \cite{Re:Koyluoglu}. Thereby, we introduce a new type of codes called the ``stable" MSCR codes, where the repair data is independent of the repair groups and the sets of helper nodes. Towards it, we find the two MSCR codes proposed in \cite{Re:N. Le Scouarnec,Re:K. W. Shum} actually are not stable while we convert the MSCR code given in \cite{Re:K. W. Shum} to a stable one, which has better secrecy capacity than the original one. In addition, we utilize information theory to give some specific results on secrecy capacity.

Although we present some results on data secrecy of MSCR codes, there are still many related research questions for further exploring. First, more examples of MSCR codes and stable MSCR codes need to be further explored. Second, we need to derive the characterization of secrecy capacity in more diverse situations than considered in this paper.

\end{document}